%Paper: hep-th/9307015
%From: RZUCCHINI@BOLOGNA.INFN.IT
%Date: Fri,  2 JUL 93 13:30 GMT
%Date (revised): Mon, 5 JUL 93 17:55 GMT

%%%%%%%%%%%%%%%%%%%%%%%%%%%%%%%%%%%%%%%%%%%%%%%%%%%%%%%%%%%%%%%%%%%%%%%%%%%%%%%
%%%%%%%%%%%%%%%%%%%%%%%%%%%%%%%%%%%%%%%%%%%%%%%%%%%%%%%%%%%%%%%%%%%%%%%%%%%%%%%
%%%            DEFORMATION THEORY OF HOLOMORPHIC VECTOR BUNDLES
%%%       EXTENDED CONFORMAL SYMMETRY AND EXTENSIONS OF 2D GRAVITY
%%%%%%%%%%%%%%%%%%%%%%%%%%%%%%%%%  by  %%%%%%%%%%%%%%%%%%%%%%%%%%%%%%%%%%%%%%%%
%%%                         Roberto Zucchini
%%%%%%%%%%%%%%%%%%%%%%%%%%%%%%%%%%%%%%%%%%%%%%%%%%%%%%%%%%%%%%%%%%%%%%%%%%%%%%%
%%%                            Plain TeX
%%%%%%%%%%%%%%%%%%%%%%%%%%%%%%%%%%%%%%%%%%%%%%%%%%%%%%%%%%%%%%%%%%%%%%%%%%%%%%%
%%%%%%%%%%%%%%%%%%%%%%%%%%%%%%%%%%%%%%%%%%%%%%%%%%%%%%%%%%%%%%%%%%%%%%%%%%%%%%%
\magnification=1200
\baselineskip=.52cm plus .52mm minus .52mm
\def\ref#1{\lbrack#1\rbrack}
\font\sf=cmss10
\def\sans#1{\hbox{\sf #1}}

\def\proj{{\rm proj}\hskip 2pt}
\def\tr{{\rm tr}\hskip 1pt}
\def\dom{{\rm Dom}\hskip 1pt}
\def\ad{{\rm ad}\hskip 1pt}

\hbox to 16.5 truecm{July 1993 \hfil DFUB--93}
\hbox to 16.5 truecm{Version 1 \hfil}
\vskip1.3cm
\centerline{\bf DEFORMATION THEORY OF HOLOMORPHIC VECTOR BUNDLES}
\centerline{\bf EXTENDED CONFORMAL SYMMETRY}
\centerline{\bf AND EXTENSIONS OF 2D GRAVITY}
\vskip.9cm
\centerline{by}
\vskip.5cm
\centerline{\bf Roberto Zucchini}
\centerline{\it Dipartimento di Fisica, Universit\`a degli Studi di Bologna}
\centerline{\it V. Irnerio 46, I-40126 Bologna, Italy}
\vskip.9cm
\centerline{\bf Abstract}
\vskip.5cm
\noindent
Developing on the ideas of R. Stora and coworkers, a formulation
of two dimensional field theory endowed with extended conformal
symmetry is given, which is based on deformation theory of holomorphic
and Hermitian spaces. The geometric background consists of a vector
bundle $E$ over a closed surface $\Sigma$ endowed with a holomorphic
structure and a Hermitian structure subordinated to it. The symmetry
group is the semidirect product of the automorphism group ${\rm Aut}(E)$
of $E$ and the extended Weyl group ${\rm Weyl}(E)$ of $E$ and acts on the
holomorphic and Hermitian structures.
The extended Weyl anomaly can be shifted into an
automorphism chirally split anomaly by adding to the action a local
counterterm, as in ordinary conformal field theory.
The dependence on the scale of the metric on the fiber of $E$
is encoded in the Donaldson action, a vector bundle
generalization of the Liouville action.
The Weyl and automorphism anomaly split into two contributions
corresponding respectively to the determinant and projectivization
of $E$. The determinant part induces an effective ordinary Weyl
or diffeomorphism anomaly and the induced central charge can be
computed.
As an application, it is shown that to any $A_1$ embedding $t$ into a
simple Lie algebra $\sans g$ and any representation $R$ of $\sans g$ one
can naturally associate a flat vector bundle $DS(t,R)$
on $\Sigma$. It is further shown
that there is a deformation of the holomorphic structure of such bundle
whose parameter fields are generalized Beltrami differentials of the
type appearing in light cone $W$ geometry and that the projective part of
the automorphism anomaly reduces to the standard $W$ anomaly in the large
central charge limit. A connection between the Donaldson action and
Toda field theory is also observed.
\vfill\eject
\vskip.6cm
\item{1.} {\bf Introduction.}
\vskip.4cm
\par
During the last few years a large body of literature has been devoted to
the study of non critical string theory and $2D$ quantum gravity.
Soon after the original discovery of a ${\rm SL}(2,{\bf R})$ symmetry in
ordinary induced $2D$ gravity \ref{1-2}, it was found that
$2D$ gravity emerged from a constrained gauged ${\rm SL}(2,{\bf R})$
WZWN model \ref{3-4}. These results suggested that other models of
induced $2D$ gravity may arise from $2D$ field theories endowed
with an extended conformal symmetry
(see ref. \ref{5} for a comprehensive and updated review).
Such trend has resulted in a wide study of $W$ strings
\ref{6-8} and $W$ gravity \ref{9-13}.

A sound formulation of $W$ gravity requires a prior understanding
of the underlying geometry and its symmetries and a systematic study
of anomalies \ref{14-16}. While considerable progress has been
toward a satisfactory formulation of $W$ geometry both in the light cone
gauge \ref{17-20} and in the covariant gauge \ref{21-23},
there still are a number of field theoretic issues which call for further
investigation, such as locality, the proper generalization of the
holomorphic factorization theorem \ref{24-26} and quantization
and gauge fixing of the geometrical fields.

In this paper, I shall attempt a formulation of a two dimensional field
theory endowed with an extended conformal symmetry having in mind
appplications to extended $2D$ gravity. The approach adopted, initiated
by R. Stora and coworkers \ref{26}, is based on deformation theory of
holomorphic and Hermitian structures of complex spaces and is summarized
below.

In ordinary conformal field theory, the geometric background consists
of a world sheet $\Sigma$ endowed with a holomorphic structure $\sans a$.
Conformal fields are sections of tensor powers of the $\sans a$--holomorphic
canonical $1$--cocycle $k$ of $\Sigma$. Conformal invariance consists
precisely in the independence of physics from coordinate choices in
$\sans a$. The holomorphic structure $\sans a$ may be `deformed' yielding
a new holomorphic structure ${\sans a}'$. Deformations are parametrized
in one-to-one fashion by the Beltrami field $\mu$. The classical action
$S_{\rm cl}$ thus depends on $\mu$ and $\bar\mu$.
The response of the
system under deformation is given by the holomorphic energy-momentum
tensor $T_{\rm cl}=\delta S_{\rm cl}/\delta\mu$ and its complex conjugate.

The consistent quantization of a conformal field theory uses the
$\zeta$ function scheme, which requires the introduction of a
metric $h$ on $\Sigma$ that is Hermitian with respect the
holomorphic structure $\sans a$. Metrics can be parametrized in terms
of the Liouville field $\phi$. This introduces a geometric degree of
freedom which is extraneous to the conformal geometry of the classical
theory. In the quantum theory, one may deform both $\sans a$ and $h$.
and so the effective action $I$ will depend explicitly on $\mu$,
$\bar\mu$ and $\phi$. $\mu$, $\bar\mu$ and $\phi$ couple to the quantum
holomorphic energy momentum tensor $T_{\rm qu}$ and its complex conjugate
and to the trace of the energy momentum tensor $T_{\rm trace}
=\delta I/\delta\phi$, respectively.

The symmetry group of ordinary conformal field theory consists
in the semidirect product of the diffeomorphism group ${\rm Diff}(\Sigma)$
of $\Sigma$ and the Weyl group ${\rm Weyl}(\Sigma)$ of $\Sigma$.
The group acts naturally on the holomorphic structures and
on the Hermitian metrics subordinated or, equivalently, on $\mu$ and $\phi$.
The $\zeta$ function scheme provides an diffeomorphism invariant but
Weyl anomalous renormalization of the effective action $I$. The Weyl
anomaly can be shifted into a diffeomorphism chirally split
anomaly by adding to $I$ a local counterterm depending on
$\mu$, $\bar\mu$ and $\phi$ \ref{27-28}. The modified effective action is
independent from $\phi$ and thus depends only on the background
conformal geometry as in the classical case.
See sects. 2, 3 and 4 for a brief review of the above topics.

The formulation proposed of extended conformal field theory
parallels as much as possible that of ordinary conformal field theory
just outlined. $\Sigma$ is extended by attaching at each point an
internal vector space. Hence, the geometric background consists of a vector
bundle $E$ endowed with a holomorphic structure $\sans A$.
Since $\sans A$ induces a holomorphic structure
$\sans a$ on the base $\Sigma$, the holomorphic geometry of $\Sigma$
is contained in that of $E$. The extended conformal fields are conformal
fields carrying internal degrees of freedom corresponding to the fiber of
$E$. Such fields are sections of an $\sans a$--holomorphic $1$--cocycle
of the form $k^{\otimes m}\otimes L$, where $L$ is the
$\sans a$--holomorphic matrix $1$--cocycle associated to
trivialization changes in $\sans A$. Extended conformal
invariance consists in the independence from trivializations
choices in $\sans A$. The holomorphic structure $\sans A$ can be deformed
into a new holomorphic structure ${\sans A}'$. The deformations are
parametrized in one-to-one fashion by the Beltrami field $\mu$ and
a $(0,1)$ gauge field $A^*$. The classical action $S_{\rm cl}$
thus depends on $\mu$, $\bar\mu$, $A^*$ and $A^{*\dagger}$.
The response of the system under
deformation is given by the holomorphic energy momentum tensor $T_{\rm cl}$
and its complex conjugate and by a holomorphic current
$J_{\rm cl}=\delta S_{\rm cl}/\delta A^*$ and its Hermitian conjugate.

The consistent quantization of an extended conformal field theory uses
the $\zeta$ function scheme, as in the standard case. This
requires the introduction of a Hermitian structure $(h^\circ,H)$ on $E$
subordinated to $\sans A$, where
$h^\circ$ is a Hermitian metric of $\Sigma$ with respect to $\sans a$ and
$h$ is a Hermitian metric on the fibers of $E$. Metrics can be
parametrized in terms of the Liouville field $\phi^\circ$ and the
Donaldson field $\Phi$. This introduces geometric degrees of freedom
extraneous to the extended conformal geometry of the classical
theory. In the quantum theory, one may deform both $\sans A$ and
$(h^\circ,H)$ and so $I$ will depend explicitly on both $\mu$, $A^*$,
$\phi^\circ$ and $\Phi$. $\mu$, $\bar\mu$ and $\phi^\circ$ couple to the
quantum holomorphic energy momentum tensor $T_{\rm qu}$ and its complex
conjugate and to the trace energy momentum tensor $T^\circ{}_{\rm trace}$.
$A^*$, $A^{*\dagger}$ and $\Phi$ couple similarly to
the quantum current $J_{\rm qu}$ and its Hermitian conjugate
and to a further current $J_{\rm an}$.

In extended conformal field theory, the symmetry group is properly
the semidirect product of the automorphism group ${\rm Aut}(E)$ of $E$
and the extended Weyl group ${\rm Weyl}(E)$ of $E$. The symmetry
group acts on the holomorphic and Hermitian structures
and hence on the geometrical fields $\mu$, $A^*$, $\phi^\circ$ and $\Phi$.
The $\zeta$ function scheme provides an automorphism invariant but extended
Weyl anomalous renormalization of the effective action of extended conformal
field theory. The extended Weyl anomaly can be shifted into an
automorphism chirally split anomaly by adding to the action a local
counterterm depending on $\mu$, $\bar\mu$, $A^*$, $A^{*\dagger}$,
$\phi^\circ$ and $\Phi$ \ref{26}. It can be seen that both
the Weyl and automorphism anomaly split into two contributions
corresponding respectively to the determinant and projectivization
of $E$. The determinant part induces an effective ordinary Weyl
or diffeomorphism anomaly and the induced central charge can be
computed. See sects. 5, 6, 7 and 8 for a discussion of these topics.

The dependence on the Donaldson field $\Phi$ is encoded in
the Donaldson action \ref{29}, a vector bundle generalization of the
Liouville action formally similar to a non compact gauged WZWN action
in which the WZ field is $\exp\Phi$ and the gauge fields are $A^*$ and
$A^{*\dagger}$. I believe that the Donaldson action may play an important
role in the formulation of models of $2D$ quantum gravity a la
David-Distler-Kaway \ref{30-31}. Integrating the gauge fields may also yield
effective actions for the Donaldson field $\Phi$ with black hole
properties \ref{32}.

To make contact with extended gravity, it is necessary to consider a
particular choice of the bundle $E$ and to restrict oneself to special
deformations of its holomorphic structure and special Hermitian
structures. The construction relies
on $A_1$ embeddings into simple Lie algebras, in the same spirit
as refs. \ref{33-34}. To any $A_1$ embedding $t$ into a simple
Lie algebra $\sans g$ and any representation $R$ of $\sans g$,
there is naturally associated a flat unstable holomorphic vector bundle
$DS(t,R)$ on any given Riemann surface. $E$ is the tensor product of a
tensor power of the canonical line bundle and $DS(t,R)$.
There are deformations of the holomorphic structure of $E$
whose parameter fields are the Beltrami differential $\mu$ and
generalized Beltrami differentials $\nu_\eta$ of the type as those
appearing in light cone $W$ geometry, one for each of the representations
of $A_1$ contained in the adjoint representation of $\sans g$.
The field space formed by the Beltrami differential $\mu$ and the
generalized Beltrami differentials $\nu_\eta$ possesses a $W$
symmetry. The symmetry is non local, but its action on the parameter
fields translates into an action on the geometrical fields $\mu$ and $A^*$
via automorphisms. In this sense, the non local $W$ symmetry is embedded
into a local one. Further, the projective part of the automorphism anomaly
reduces to the standard $W$ anomaly in the large central charge limit.
There is also a class of Hermitian structures of $E$ parametrized by
the Liouville field $\phi^\circ$ and by Toda fields $\phi_\alpha$,
one for each of the simple roots of $\sans g$. When expressed in terms of
such fields, the Donaldson action assumes a form closely related to
Toda field theory.
See sect. 9, for an illustration of these applications.
\vskip.6cm
\item{2.} {\bf Holomorphic and Hermitian geometry of higher genus surfaces.}
\vskip.4cm
\par
In this section, I shall provide a brief account of the basic notions of
holomorphic and Hermitian geometry of a surface (see refs. \ref{35-36}
for background material). The geometrical framework illustrated below
will be the starting point for the generalizations of sect. 5. and the
results summarized here will be repeatedly invoked in later sections.

Let $\Sigma$ be a connected compact smooth oriented surface of genus
$\ell\geq 2$. A holomorphic structure $\sans a$ on $\Sigma$ is a
maximal atlas of local coordinates $\{z_a\}$ with holomorphic coordinate
changes contained in the oriented differentiable structure of $\Sigma$
\footnote{${}^1$}{In this paper, I shall denote coordinate
patches and bundle trivializations by early lower case Latin subscripts.
When no confusion is possible, I shall suppress such labels to simplify
the notation.}. The $\sans a$--holomorphic canonical
$1$--cocycle $k$ of $\Sigma$ is defined by $k_{ab}=\partial_a
z_b$ for $\dom z_a\cap\dom z_b\not=\emptyset$,
where $\partial_a=\partial/\partial z_a$.

In conformal field theory, one compares several holomorphic structures
$\sans a$, ${\sans a}'$, ${\sans a}''$ etc. \footnote{${}^2$}{
I shall adopt the convention of attaching a corresponding
number of primes to all objects referring to each of them.}.
Chosen a reference holomorphic structure $\sans a$,
any other holomorphic structure ${\sans a}'$ can be viewed as
obtained by a continuous deformation of $\sans a$. Since
$\sans a$ and  ${\sans a}'$ belong to
the same oriented differentiable structure of $\Sigma$
$$\partial_az'{}_{a'}\bar\partial_a\bar z'{}_{a'}
-\bar\partial_az'{}_{a'}\partial_a\bar z'{}_{a'}>0.\eqno(2.1)$$

Each deformation ${\sans a}'$ is characterized by a
field $\lambda=\lambda({\sans a}')$ locally given by
$$\lambda_{aa'}=\partial_az'{}_{a'},\eqno(2.2)$$
for any two overlapping coordinates $z_a$ and $z'{}_{a'}$
from $\sans a$ and ${\sans a}'$ \ref{37}.
$\lambda$ belongs to $S(\Sigma,k\otimes k'^{-1})$
\footnote{${}^3$}{I shall denote by
$S(\Sigma, Z)$ the space of smooth section of a smooth $1$--cocycle $Z$.}.
It follows from $(2.1)$ that $\lambda$ is nowhere
vanishing. As will appear in due course, $\lambda$ is a deformation
intertwiner.

To each deformation ${\sans a}'$, one can canonically
associate a field $\mu=\mu({\sans a}')$ locally defined by \ref{37}
$$\mu_a=\bar\partial_az'{}_{a'}\lambda_{aa'}{}^{-1}.\eqno(2.3)$$
It is easily verified that $\mu_a$ is independent from the coordinate
$z'{}_{a'}$ of ${\sans a}'$ chosen, as suggested by the notation. Further,
$\mu$ belongs to $S(\Sigma,k^{-1}\otimes\bar k)$ and satisfies the bound
$$\sup_\Sigma |\mu|<1,\eqno(2.4)$$
which follows directly from $(2.1)$.
A field $\mu$ with such properties is called
a Beltrami differential. The map ${\sans a}'\rightarrow\mu({\sans a}')$
can be inverted. To any Beltrami differential $\mu$, one can associate
canonically a deformation ${\sans a}'={\sans a}(\mu)$ of the holomorphic
structure whose generic coordinate $z'{}_{a'}$ is a local solution of the
Beltrami equation
$$(\bar\partial_a-\mu_a\partial_a)z'{}_{a'}=0\eqno(2.5)$$
subject to the local invertibility condition $(2.1)$. The Beltrami
differential $\mu$ associated to ${\sans a}'$ is precisely the field $\mu$
appearing in $(2.5)$, as is evident from comparing $(2.3)$ and $(2.5)$.
Hence, the set ${\rm Beltr}(\Sigma)$ of all Beltrami differentials $\mu$
parametrizes in one-to-one fashion the family of all deformations
${\sans a}'$. Note that ${\sans a}'={\sans a}$ for $\mu=0$.
${\rm Beltr}(\Sigma)$ is an infinite dimensional holomorphic manifold.
$\lambda$ is a non local holomorphic functional of $\mu$.

By using $(2.2)$ and $(2.5)$, one can easily verify that \ref{37}
$$dz'{}_{a'}=(dz_a+\mu_ad\bar z_a)\lambda_{aa'},\eqno(2.6)$$
$$\bar\partial'{}_{a'}={1\over (1-\bar\mu_a\mu_a)\bar\lambda_{aa'}}
(\bar\partial_a-\mu_a\partial_a).\eqno(2.7)$$
These formulae are often employed in calculations.

For any holomorphic structure $\sans a$, there exists an
$\sans a$--holomorphic $1$-cocycle $k^{\otimes{1\over 2}}$ such
that $(k^{\otimes{1\over 2}})^{\otimes 2}=k$ \ref{35}. In fact, there are
$2^{2\ell}$ choices of $k^{\otimes{1\over 2}}$, each of which corresponds to
a spinor structure of $\Sigma$. Below, I shall work with a fixed but
otherwise arbitrary choice.

Given a deformation ${\sans a}'$ of $\sans a$, define
$$\lambda^{\otimes{1\over 2}}{}_{aa'}=\lambda_{aa'}{}^{1\over 2},\eqno(2.8)$$
with some choice of the branch of the square root for which the right hand
side is smooth as a local function in $\Sigma$ and holomorphic as a functional
of $\mu$. Under a change of coordinates in $\sans a$, one has
$\lambda^{\otimes{1\over 2}}{}_{aa'}
=\eta^{a'}{}_{ab}k^{\otimes{1\over 2}}{}_{ab}
\lambda^{\otimes{1\over 2}}{}_{ba'}$, where $\eta^{a'}{}_{ab}=\pm 1$.
$\eta^{a'}$ is a trivial ${\bf Z}_2$--valued $1$--cocycle over
$\dom z'{}_{a'}$, since the latter can be assumed simply connected without any
loss of generality. Thus, the branch of the square root in $(2.8)$ can be
chosen in such a way that $\eta^{a'}{}_{ab}=1$ identically.
The square root is fixed by demanding that
$\lambda^{\otimes{1\over 2}}{}_{ab'}=k^{\otimes{1\over 2}}{}_{ab}$ for
$\mu=0$. It follows from here that $k'^{\otimes{1\over 2}}{}_{a'b'}=
\lambda^{\otimes-{1\over 2}}{}_{aa'}k^{\otimes{1\over 2}}{}_{ab}
\lambda^{\otimes{1\over 2}}{}_{bb'}$ is a tensor square root of $k'$
depending holomorphically on the Beltrami field $\mu$.

A conformal field $\psi$ of weights $m,\bar m\in {\bf Z}/2$
with respect to the holomorphic structure $\sans a$
is any element of $S(\Sigma,k^{\otimes m}\otimes\bar k^{\otimes \bar m})$.
The intertwiner
$\lambda^{\otimes{1\over 2}}$ induces a linear isomorphism of the spaces
$S(\Sigma,k^{\otimes m}\otimes\bar k{\vphantom k}^{\otimes \bar m})$ and
$S(\Sigma,k'^{\otimes m}\otimes\bar k'{\vphantom k}^{\otimes \bar m})$ of
conformal fields of weights $m,\bar m$ associated to the holomorphic
structures $\sans a$ and ${\sans a}'$, respectively \ref{37}. Such
isomorphism is given by
$$\psi'{}_{a'}=\lambda^{\otimes-m}{}_{aa'}
\bar\lambda^{\otimes-\bar m}{}_{aa'}\psi_a,\eqno(2.9)$$
for $\psi$ in $S(\Sigma,k^{\otimes m}\otimes\bar k{\vphantom k}^{\otimes
\bar m})$.

A Hermitian metric $h$ on $\Sigma$ subordinated to a given holomorphic
structure ${\sans a}$ is an element of $S(\Sigma,k\otimes\bar k)$ such that
$h_a$ takes only strictly positive values in its domain.

Every Hermitian metric $h$ can be represented as
$$h_a=\exp\phi_a g_a,\eqno(2.10)$$
where $g$ is a fixed fiducial Hermitian metric and $\phi$ is a real
valued field of $S(\Sigma,1)$, called the Liouville field.
One may view $h$ as a deformation of $g$ and $\phi$ as a field
parametrizing such deformation.

To a metric $h$,
there is associated canonically a $(1,0)$ affine connection $\gamma_h$
compatible with $h$ given locally by
$$\gamma_{ha}=\partial_a\ln h_a.\eqno(2.11)$$
The Ricci scalar $R_h$ of $h$ is given locally by
$$R_h=-2h_a{}^{-1}\bar\partial_a\partial_a\ln h_a.\eqno(2.12)$$
The covariant derivative associated to $\gamma_h$ will be denoted by
$\partial_h$.

A deformation ${\sans a}'$ of the holomorphic structure $\sans a$ induces a
deformation $h'$ of any given Hermitian metric $h$ via $(2.9)$. Namely,
$h'{}_{a'}=h_a|\lambda_{aa'}|^{-2}$. In the Liouville parametrization, one has
$h'=\exp\phi g'$. Hence, the Liouville field is invariant under deformations.
\vskip.6cm
\item{3.} {\bf The diffeomorphism and Weyl groups and the Slavnov operator.}
\vskip.4cm
\par
The next step is naturally the analysis of the underlying symmetry of the
geometrical setting described in the previous section.  As well-known,
the symmetry group is the diffeomorphism group of the
surface $\Sigma$ extended by the Weyl group of $\Sigma$.
Its properties, which will be recalled repeatedly in later sections,
are summarized below.

I shall restrict to the group ${\rm Diff}_c(\Sigma)$
of orientation preserving
diffeomorphisms of $\Sigma$ homotopically connected to the identity
${\rm id}_\Sigma$ in the $C^\infty$ topology. ${\rm Diff}_c(\Sigma)$
acts on the family of deformations ${\sans a}'$ of a reference
holomorphic structure $\sans a$ of $\Sigma$. In fact, for any
diffeomorphism $f\in{\rm Diff}_c(\Sigma)$, the maps
$$z''{}_{a''}=z'{}_{b'}\circ f\eqno(3.1)$$
form a new holomorphic structure ${\sans a}''$, the pull-back
$f^*{\sans a}'$ of ${\sans a}'$ by $f$ \ref{37}.
The pull-back action of ${\rm Diff}_c(\Sigma)$ on the deformations
${\sans a}'$ induces a right action of ${\rm Diff}_c(\Sigma)$
on the corresponding intertwiner fields $\lambda$ and Beltrami differentials
$\mu$ by the relations $f^*\lambda({\sans a}')=\lambda(f^*{\sans a}')$ and
$f^*\mu({\sans a}')=\mu(f^*{\sans a}')$.
{}From $(3.1)$ and $(2.2)$ and $(2.3)$, one finds easily that \ref{37}
$$(f^*\lambda)_{aa''}=(\partial_af_b+\mu_b\circ f\partial_a\bar f_b)
\lambda_{bb'}\circ f,\eqno(3.2)$$
$$(f^*\mu)_a={\bar\partial_af_b+\mu_b\circ f\bar\partial_a\bar f_b\over
\partial_af_b+\mu_b\circ f\partial_a\bar f_b}.\eqno(3.3)$$

${\rm Diff}_c(\Sigma)$ acts on the space
$S(\Sigma,k^{\otimes m}\otimes\bar k{\vphantom k}^{\otimes \bar m})$
of conformal fields $\psi$ of weights $m,\bar m$. The action is such
that one has
$$(f^*\psi)''{}_{a''}=\psi'_{b'}\circ f,\eqno(3.4)$$
where ${\sans a}''=f^*{\sans a}'$ \ref{37}. This is consistent since, as
can easily be shown,
$k''^{\otimes{1\over 2}}{}_{a''c''}=k'^{\otimes{1\over 2}}{}_{b'd'}\circ f$,
where ${\sans a}''=f^*{\sans a}'$ and $z''{}_{a''}, z''{}_{c''}$
are related to $z'{}_{b'},z'{}_{d'}$ by $(3.1)$, respectively.
{}From $(2.8)$, $(2.9)$ and $(3.2)$, one sees that
$$f^*\psi_a=\varpi_{ab}(f;\mu)^{2m}\bar\varpi_{ab}(f;\mu)^{2\bar m}\psi_b
\circ f,\eqno(3.5)$$
where
$$\varpi_{ab}(f;\mu)=(\partial_af_b+\mu_b\circ f
\partial_a\bar f_b)^{1\over 2}.\eqno(3.6)$$
The branch of the square root is defined uniquely by taking the square root
of both sides of $(3.2)$ and demanding that
$(f^*\lambda^{\otimes{1\over 2}})_{aa''}=\varpi_{ab}(f;\mu)
\lambda^{\otimes{1\over 2}}{}_{bb'}\circ f$, where the square root
$\lambda^{\otimes{1\over 2}}$ has been defined in sect. 2.
Using the remark at the beginning of this paragraph, it is easy to show
that the branch cannot depend on the coordinates $z''{}_{a''}$
and $z'{}_{b'}$.

When Hermitian structures are envisaged, the symmetry group is
enlarged to the semidirect product ${\rm Diff}_c(\Sigma)\times
{\rm Weyl}(\Sigma)$, where ${\rm Weyl}(\Sigma)\cong\exp S(\Sigma,1)$ is
the Weyl group of $\Sigma$, the group multiplication being defined by
$$(f_1,\upsilon_1)(f_2,\upsilon_2)=(f_1\circ f_2,
\upsilon_1f_1{}^{-1*}\upsilon_2).\eqno(3.7)$$

The action of a combined diffeomorphism and Weyl rescaling
$(f,\upsilon)$ on the geometric fields $\lambda$ and $\mu$
and on conformal fields reduces to that of $f$ given above. For a
Hermitian metric $h$, one has instead
$$f^{*\upsilon}h=f^*\big(\upsilon^{-1}h\bar\upsilon^{-1}\big).\eqno(3.8)$$
This can easily be transcribed into an action on the Liouville field $\phi$
(cf. eq. $(2.10)$).

In field-theoretic analyses, one is mostly interested in the infinitesimal
form of the symmetry. One is thus lead to considering the Lie algebra
${\rm Lie}\hskip1pt\big({\rm Diff}_c(\Sigma)\times{\rm Weyl}(\Sigma)\big)$
of ${\rm Diff}_c(\Sigma)\times{\rm Weyl}(\Sigma)$.
Writing formally $f={\rm id}_\Sigma+\xi$ and $\upsilon=1-\theta/2$ in
$(3.7)$, it is easy to verify ${\rm Lie}\hskip1pt\big({\rm Diff}_c(\Sigma)
\times{\rm Weyl}(\Sigma)\big)$ is isomorphic to the semidirect sum
$S(\Sigma,k^{-1})\oplus S(\Sigma,1)$ with the Lie brackets
$$[\xi_1\oplus\theta_1,\xi_2\oplus\theta_2]=\big((\xi_1\partial+\bar\xi_1
\bar\partial)\xi_2-(\xi_2\partial+\bar\xi_2\bar\partial)\xi_1\big)\oplus
\big((\xi_1\partial+\bar\xi_1\bar\partial)\theta_2-(\xi_2\partial+
\bar\xi_2\bar\partial)\theta_1\big).\eqno(3.9)$$

To express the infinitesimal action of the symmetry group
${\rm Diff}_c(\Sigma)\times{\rm Weyl}(\Sigma)$ on field functionals,
one introduces the exterior algebra
$\bigwedge^*{\rm Lie}\hskip1pt\big({\rm Diff}_c(\Sigma)\times
{\rm Weyl}(\Sigma)\big){\vphantom )}^\vee$ of
${\rm Lie}\hskip1pt\big({\rm Diff}_c(\Sigma)\times
{\rm Weyl}(\Sigma)\big)$. The algebra is generated by the diffeomorphism
ghost $c$ and the Weyl ghost $w$. These are the sections of $k^{-1}\otimes
\bigwedge^1{\rm Lie}\hskip1pt\big({\rm Diff}_c(\Sigma)\times
{\rm Weyl}(\Sigma)\big){\vphantom )}^\vee$ and $1\otimes\bigwedge^1{\rm Lie}
\hskip1pt\big({\rm Diff}_c(\Sigma)\times{\rm Weyl}(\Sigma)\big)
{\vphantom )}^\vee$ defined by $\langle c(p),\xi\oplus\theta\rangle=\xi(p)$
and $\langle w(p),\xi\oplus\theta\rangle=\theta(p)$ for $p\in\Sigma$,
respectively. By a standard construction \ref{37}, the linearization of
the action of the symmetry group on the relevant space $\cal F$ of field
functionals at the identity, defines a coboundary operator
$s$ on $\bigwedge^*{\rm Lie}\big(\hskip1pt{\rm Diff}_c(\Sigma)\times{\rm Weyl}
(\Sigma)\big){\vphantom )}^\vee\otimes{\cal F}$, called the Slavnov operator.
$s^2=0$, that is $s$ is nilpotent. From $(3.9)$, it is not difficult to read
off the structure equations
$$sc=(c\partial+\bar c\bar\partial)c,\eqno(3.10)$$
$$sw=(c\partial+\bar c\bar\partial)w.\eqno(3.11)$$
For a given Beltrami differential $\mu$ in
${\rm Beltr}(\Sigma)$, the relevant combination of ghost fields is
\ref{37}
$$C=c+\mu\bar c.\eqno(3.12)$$
$C$ exhibits a distinguished complex structure of
${\rm Lie}\hskip1pt{\rm Diff}_c(\Sigma)$. The expressions of
$s\lambda$ and $s\mu$ can be easily obtained from $(3.2)$, using
$(2.5)$, and $(3.3)$:
$$s\lambda_{a'}=\partial(C\lambda_{a'}),\eqno(3.13)$$
$$s\mu=\big(\bar\partial-\mu\partial+(\partial\mu)\big)C.\eqno(3.14)$$
By combining $(3.10)$ and $(3.14)$, one finds further that $C$ obeys the
structure equation
$$sC=C\partial C.\eqno(3.15)$$
For any conformal field $\psi$ of weights $m,\bar m$, one has
$$s\psi'=(c\partial+\bar c\bar\partial)\psi',\eqno(3.16)$$
$$s\psi=(c\partial+\bar c\bar\partial)\psi+\big(m(\partial c+\mu\partial\bar c)
+\bar m(\bar\partial\bar c+\bar\mu\bar\partial c)\big)\psi,\eqno(3.17)$$
as follows easily from $(3.4)$--$(3.5)$ upon linearization.
As to the Liouville field, one has
$$s\phi=(c\partial+\bar c\bar\partial)\phi+(\partial_g+\bar\mu\bar\partial)c
+w/2+(\bar\partial_g+\mu\partial)\bar c+\bar w/2,
\quad sg=0.\eqno(3.18)$$
Note that the Liouville field $\phi$ does not transform as a conformal field
of weights $0,0$.
\vskip.6cm
\item{4.} {\bf Conformal field theory: Weyl versus diffeomorphism anomaly.}
\vskip.4cm
\par
Consider a classical conformally invariant local field theory on a surface
$\Sigma$ endowed with a holomorphic structure $\sans a$. Its quantization is
carried out by means of the $\zeta$ function renormalization scheme. Such
method requires the introduction of a Hermitian metric $h$ with respect
to $\sans a$ to properly define the adjoint of the relevant differential
operators. The effective action $I_{\rm diff}(h)$ will thus depend on
$h$. One may fix a reference holomorphic structure $\sans a$ and a Hermitian
metric $h$ subordinated to $\sans a$ and consider any deformation ${\sans a}'$
of $\sans a$ and the associated deformation $h'$ of the metric $h$. To this
there corresponds an effective action $I'{}_{\rm diff}(h')$. The symmetry
group ${\rm Diff}_c(\Sigma)\times{\rm Weyl}(\Sigma)$ acts on ${\sans a}'$
and $h'$ as described in sect. 3 and thus on $I'{}_{\rm diff}(h')$.
The $\zeta$ function renormalization scheme is diffeomorphism invariant,
that is $I'{}_{\rm diff}(h')$ is invariant under the action of
${\rm Diff}_c(\Sigma)$. But $I'{}_{\rm diff}(h')$ will depend in general
on the scale of $h'$ and not simply on the holomorphic structure
${\sans a}'$. A Weyl anomaly is produced in this way. The form of such
anomaly is universal:
$$sI'{}_{\rm diff}(h')=\kappa{\cal A}'{}_{\rm conf}(w_1;h'),\eqno(4.1)$$
where $\kappa$ is the central charge of the model under consideration and
$$\eqalignno{&{\cal A}'{}_{\rm conf}(w_1;h')=
{\cal A}_{\rm conf}(w_1,\phi,\mu,\bar\mu;g)=
{-1\over 12\pi}\int_\Sigma{d\bar z'\wedge dz'\over 2i}w_1
\partial'\bar\partial'\ln h'
={-1\over 12\pi}\int_\Sigma{d\bar z\wedge dz\over 2i}w_1&\cr&\bigg\{
\partial\bar\partial\ln g+\partial\bar\partial\phi
-\big(\partial-\bar\mu\bar\partial-(\bar\partial\bar\mu)\big)
{\partial_g\mu+\partial\phi\mu\over 1-\mu\bar\mu}
-\big(\bar\partial-\mu\partial-(\partial\mu)\big)
{\bar\partial_g\bar\mu+\bar\partial\phi\bar\mu
\over 1-\mu\bar\mu}\bigg\},&(4.2)\cr}$$
$$w_1=(w+\bar w)/2.\eqno(4.3)$$
Here, the dependence on the deformation ${\sans a}'$ and the scale of
the metric is given explicitly in terms of the Beltrami field $\mu$
and the Liouville field $\phi$ relative to a reference metric $g$
to make manifest the locality of ${\cal A}'{}_{\rm conf}$.

The Weyl anomaly can be eliminated by either $i$)
constraining the field content of the model so that the total
central charge vanishes, as in the case of string theory \ref{38}, or $ii$)
subtracting from the effective action a suitable local counterterm that
absorbs the Weyl anomaly at the cost of creating a diffeomorphism anomaly
\ref{27-28}. The form of such counterterm is also universal. Up to an
overall factor $\kappa$, the counterterm is a sum of two contributions
The first contribution is just the Liouville action $S'{}_L(\phi;g')$
(with no cosmological term)
expressing the dependence on the scale of the metric $h$
$$\eqalignno{&S'{}_L(\phi;g')=S_L(\phi,\mu,\bar\mu;g)
={-1\over 12\pi}\int_\Sigma{d\bar z'\wedge dz'\over 2i}\bigg({1\over2}
\partial'\phi\bar\partial'\phi-\partial'\bar\partial'\ln g'\phi\bigg)
&\cr&={-1\over 12\pi}\int_\Sigma{d\bar z\wedge dz\over 2i}
\bigg\{{1\over 2(1-\mu\bar\mu)}(\partial-\bar\mu\bar\partial)\phi
(\bar\partial-\mu\partial)\phi
-\bigg\lbrack\partial\bar\partial\ln g
-\big(\partial-\bar\mu\bar\partial-(\bar\partial\bar\mu)\big)
{\partial_g\mu\over 1-\mu\bar\mu}
&\cr&-\big(\bar\partial-\mu\partial-(\partial\mu)\big)
{\bar\partial_g\bar\mu\over 1-\mu\bar\mu}\bigg\rbrack\phi\bigg\}.&(4.4)\cr}$$
The second contribution is the Verlinde-Knecht-Lazzarini-Thuillier action,
worked out successively in refs. \ref{27-28}, and is given by
$$\eqalignno{&S_{VKLT}(\mu,\bar\mu;g,{\cal R},\bar{\cal R})
={1\over 12\pi}\int_\Sigma{d\bar z\wedge dz\over 2i}\bigg\{\mu({\cal R}
-r_g)+\bar\mu(\bar{\cal R}-\bar r_g)
-{1\over(1-\mu\bar\mu)}\bigg\lbrack\partial_g\mu\bar\partial_g\bar\mu
&\cr&-{1\over 2}\bar\mu\big(\partial_g\mu\big)^2
-{1\over 2}\mu\big(\bar\partial_g\bar\mu\big)^2\bigg\rbrack
\bigg\},&(4.5a)\cr}$$
where $r_g$ is the projective connection
\footnote{${}^4$}{A projective connection $R$ is a collection $\{R_a\}$ of
local smooth maps gluing as $R_b=k_{ba}(R_a-\{z_b,z_a\})$, where
$\{f,\zeta\}=-2(\partial_\zeta f)^{1\over 2}\partial_\zeta{}^2
(\partial_\zeta f)^{-{1\over 2}}$ is the Schwarzian derivative.
$\sans a$--holomorphic projective connections are known to exist \ref{35}.}
$$r_g=\partial\gamma_g-{1\over 2}\gamma_g{}^2\eqno(4.5b)$$
and ${\cal R}$ is a holomorphic projective connection in the reference
holomorphic structure $\sans a$ satisfying
$$s{\cal R}=0.\eqno(4.6)$$
${\cal R}$ ensures the correct conformal covariance of the integrand
in the right hand side of eq. $(4.5a)$.
The counterterm is thus
$$\Delta I_{\rm cft}(\phi,\mu,\bar\mu;g,{\cal R},\bar{\cal R})
=S_L(\phi,\mu,\bar\mu;g)+S_{VKLT}(\mu,\bar\mu;g,{\cal R},\bar{\cal R}).
\eqno(4.7)$$
The Weyl invariant effective action
$I_{\rm conf}$ is obtained by adding the counterterm to
$I_{\rm diff}$ \ref{28}:
$$I_{\rm conf}(\mu,\bar\mu;{\cal R},\bar{\cal R})
=I'{}_{\rm diff}(h')
+\kappa\Delta I_{\rm cft}(\phi,\mu,\bar\mu;g,{\cal R},\bar{\cal R}).
\eqno(4.8)$$
$I_{\rm conf}$ depends on the
Beltrami fields $\mu$, $\bar\mu$ and the background projective connections
${\cal R}$ and $\bar{\cal R}$. Crucially, Weyl invariance ensures that no
dependence on $g$ occurs. $I_{\rm conf}$ obeys the Ward identity
$$sI_{\rm conf}(\mu,\bar\mu;{\cal R},\bar{\cal R})=
\kappa{\cal A}_{\rm diff}(C,\mu;{\cal R})
+\kappa\overline{{\cal A}_{\rm diff}(C,\mu;{\cal R})},\eqno(4.9)$$
where
$${\cal A}_{\rm diff}(C,\mu;{\cal R})
={1\over 12\pi}\int_\Sigma{d\bar z\wedge dz\over 2i}C\big(\partial^3+
2{\cal R}\partial+(\partial{\cal R})\big)\mu\eqno(4.10)$$
\ref{28}. The Weyl anomaly has been traded for a diffeomorphism
anomaly whose strength is again measured by the central charge $\kappa$.
The salient feature of the diffeomorphism anomaly is that it is chirally
split:
it is the sum of two contributions each of which depends on only one of the
pairs $(C,\mu)$ and $(\bar C,\bar\mu)$. This fact is intimately related to
the holomorphic factorization property of $I_{\rm conf}$,
that is the chiral splitting of $I_{\rm conf}$ itself:
$I_{\rm conf}(\mu,\bar\mu;{\cal R},\bar{\cal R})=
I_P(\mu;{\cal R})+\overline{I_P(\mu;{\cal R})}$
\ref{24-25}. The functional $I_P$ is called the Polyakov action
of the model under consideration \ref{1,39}.

It is important to have in mind a standard example, the bosonic
spin $j$ $b$--$c$ system, $j\in{\bf Z}/2$. For any holomorphic structure
$\sans a$, the classical action reads
$$S(\psi^\vee,\psi)={1\over\pi}\int_{\Sigma}{d\bar z\wedge dz\over 2i}
\psi^\vee\bar\partial\psi,\eqno(4.11)$$
where $\psi$ and $\psi^\vee$ vary, respectively, in
$S(\Sigma, k^{\otimes j})$ and $S(\Sigma, k^{\otimes 1-j})$.
The central charge of this model can be computed in a variety of ways.
The well-known result is
$$\kappa=2(6j^2-6j+1).\eqno(4.12)$$
\vskip.6cm
\item{5.} {\bf Holomorphic and Hermitian geometry of complex vector bundles
over higher genus Riemann surfaces.}
\vskip.4cm
\par
In sect. 2,  I have given a brief account of the basic notions and results
of holomorphic geometry of a surface $\Sigma$ from the point of view of
deformation theory. In this section, I shall show that such geometrical
framework has a natural generalization where $\Sigma$ is replaced by
smooth complex vector bundle $E$ over $\Sigma$. The section is divided
in four subsections. In subsect. $a$, I shall indicate the basic
topological properties of $E$. Subsect. $b$ describes the holomorphic
and Hermitian geometry of $E$ in the language of deformation theory.
Subsect. $c$ analyzes the relation between the holomorphic and Hermitian
geometry of $E$ and that of its determinant and projectivization.
Finally, subsect. $d$ introduces the notion of special holomorphic and
Hermitian geometry which emerges naturally in field theoretic applications.

Before entering into the details of the discussion,
it is necessary to define the notation used and recall a
few basic concepts (see refs. \ref{36,40-41} for background material).
For any smooth fiber bundle $F$ over a smooth manifold $M$, I shall denote
the bundle projection by $\pi_F$, or simply by $\pi$ when no confusion is
possible, and the fiber of $F$ at a point $p$ of $M$ by $F_p$. An
isomorphism $T:F\rightarrow F'$ of fiber bundles is a diffeomorphism of $F$
onto $F'$ with the following properties. There exists a diffeomorphism
$f_T:M\rightarrow M'$ of their bases satisfying $\pi_{F'}\circ T=f_T\circ
\pi_F$. $T\big |_{F_p}$ is a diffeomorphism of $F_p$ onto $F_{f_T(p)}$. In
case $F$ and $F'$ are either vector or projective bundles, it is further
required that $T\big |_{F_p}$ is either a linear or a projective isomorphism,
respectively.
I shall denote by det the covariant functor from the category of smooth
complex vector bundles equipped with the isomorphisms to the category of rank
$1$ smooth vector bundles equipped with the isomorphisms. By definition, for
any complex vector bundle $X$ over $M$, $\det X$ is the rank $1$ complex vector
bundle over $M$ whose fiber $\det X_p$ at any point $p$ of $M$ is the
determinant space of the fiber $X_p$ of $X$ at $p$. For any
isomorphism $T:X\rightarrow Y$ of complex vector bundles, $\det T$ is the
isomorphism $\det X\rightarrow\det Y$ such that $\det T\big |_{\det X_p}$
is the ordinary determinant of the linear isomorphism $T\big |_{X_p}$.
I shall denote by proj the covariant functor from the category of smooth
complex vector bundles equipped with the isomorphisms to the category of smooth
complex projective vector bundles equipped with the isomorphisms. By
definition,
for any complex vector bundle $X$ over $M$, $\proj X$ is the complex
projective bundle over $M$ whose fiber $\proj X_p$ at any point $p$ of $M$
is the projective space of the fiber $X_p$ of $X$ at $p$. For any
isomorphism $T:X\rightarrow Y$ of complex vector bundles, $\proj T$ is the
isomorphism $\proj X\rightarrow\proj Y$ such that
$\proj T\big |_{\proj X_p}$ is the projective isomorphism canonically
associated to the linear isomorphism $T\big |_{X_p}$ \footnote{${}^5$}{Given
a linear isomorphism $L:V\rightarrow W$ of vector spaces, the associated
projective isomorphism $\hat L:P(V)\rightarrow P(W)$ of the corresponding
projective spaces is uniquely defined by $\hat L\circ \varrho_V=\varrho_W
\circ L$, $\varrho_X$ being the natural projection of a vector space $X$
onto its projective space.}.
\vskip.4cm
$a)$ {\it Topological properties of E.}
\vskip.4cm
The basic topological structure which will be considered here is a smooth
rank $r$ complex vector bundle $E$ over a connected compact smooth oriented
surface $\Sigma$ of genus $\ell\geq 2$.

The Chern number $c_1(\det E)$ of $\det E$ is by definition the degree $d$
of $E$. Now, it is well-known that the smooth line bundles are classified
topologically by their Chern number. Let $\epsilon$ be the smooth line bundle
of $\Sigma$ such that $c_1(\epsilon)=2(\ell-1)$. ($\epsilon$ corresponds to
the holomorphic canonical line bundle upon choosing a holomorphic structure
on $\Sigma$). There is an integer $p>0$ and a halfinteger $q$ such that
$pc_1(\det E)=qc_1(\epsilon)$, i. e. $c_1\big((\det E)^{\otimes p}\big)=
c_1(\epsilon^{\otimes q})$. One thus has a relation of the form
$$(\det E)^{\otimes p}=\epsilon^{\otimes q}.\eqno(5.1)$$
Of course, $p$ and $q$ are defined only up to multiplication by a common
positive integer number. One can however fix them by requiring that $p$ and
$2|q|$ are relative prime if $q\not=0$ and that $p=1$ if $q=0$.
The ratio
$$j=q/pr\eqno(5.2)$$
is conversely an intrinsic property of $E$. It will be called the
normalized weight of $E$. $j$ is related to the slope $s$ of the vector
bundle $E$:
$$s=d/r=2j(\ell-1).\eqno(5.3)$$
\vskip.4cm
$b)$ {\it Holomorphic and Hermitian geometry of E.}
\vskip.4cm
When viewed as a real manifold, $E$ is oriented.
A holomorphic structure $\sans A$ of $E$ is a
maximal collection $\{(z_a,u_a)\}$ of local trivializations contained in
the oriented differentiable structure of $E$ with the following
properties \ref{40-41}. $i)$ ${\sans a}=\{z_a\}$ is a holomorphic structure
of the base $\Sigma$; $ii)$ for any two overlapping trivializations, one has
$$u_b=L_{ba}\circ\pi u_a,\eqno(5.4)$$
where $L=\{L_{ba}\}$ is an $\sans a$--holomorphic
${\rm GL}(r,{\bf C})$--valued $1$--cocycle.
The holomorphic structure $\sans a$ of $\Sigma$ and the $\sans a$--holomorphic
${\rm GL}(r,{\bf C})$--valued $1$--cocycle $L$ are canonically associated to
$\sans A$.

In extended conformal field theory, one compares several holomorphic
structures $\sans A$, ${\sans A}'$, ${\sans A}''$ etc. of $E$.
The following analysis parallels closely that of sect. 2. Chosen a reference
holomorphic structure $\sans A$, any other holomorphic structure ${\sans A}'$
can be viewed as obtained from $\sans A$ by means of a continuous deformation.
Associated to ${\sans A}'$ is a deformation ${\sans a}'$ of the underlying
holomorphic structure of $\sans a$.

Each deformation ${\sans A}'$ is characterized by the
intertwiner field $\lambda$ of the underlying deformation ${\sans a}'$
defined in $(2.2)$ and by a further intertwiner field $V=V({\sans A}')$
defined as follows. The relation between the trivializing
mappings $u_a$ and $u'{}_{a'}$ of two overlapping trivializations
of $\sans A$ and ${\sans A}'$ must be linear and non singular. Thus,
it is of the form \ref{26}
$$u_a=V_{aa'}\circ\pi u'{}_{a'},\eqno(5.5)$$
where $V_{aa'}$ is some smooth ${\rm GL}(r,{\bf C})$--valued function. $V$
is by definition the collection $\{V_{aa'}\}$. From $(5.4)$ and $(5.5)$,
it appears that $V$ belongs to $S(\Sigma,L\otimes L'^\vee)$
\footnote{${}^6$}{For any smooth ${\rm GL}(r,{\bf C})$--valued $1$--cocycle
$Z$, I shall denote by $Z^\vee$ the dual cocycle
$\{Z_{ab}{}^{-1t}\}$ of $Z$ and by ${\rm End}\hskip1pt Z$
the cocycle $Z\otimes Z^\vee$.}.

To any deformation ${\sans A}'$ of the holomorphic
structure of $E$, one can canonically associate two geometrical fields.
The first one is the Beltrami differential $\mu=\mu({\sans a}')$ of the
underlying deformation ${\sans a}'$ of the holomorphic structure of
$\Sigma$ defined by $(2.3)$. The second one is defined as follows.
Let $A$ be a fixed fiducial $(1,0)$ connection of the $\sans a$--holomorphic
${\rm GL}(r,{\bf C})$--valued $1$--cocycle $L$ so that
$A_b=k_{ba}\big(L_{ba}A_a L_{ba}{}^{-1}+\partial_aL_{ba}
L_{ba}{}^{-1}\big)$. Consider the field $A^*=A^*({\sans A}')$
locally defined by \ref{26}
$$A^*{}_a=(\bar\partial_a-\mu_a\partial_a)V_{aa'}V_{aa'}{}^{-1}+\mu_aA_a.
\eqno(5.6)$$
It is easily verified that $A^*{}_a$ does not depend on the choice of the
trivialization $(z'{}_{a'},u'{}_{a'})$ of ${\sans A}'$, as suggested by the
notation, since $0=\bar\partial'{}_{b'}L'{}_{a'b'}\propto(\bar\partial_b
-\mu_b\partial_b)L'{}_{a'b'}$ by the ${\sans a}'$--holomorphy of $L'$ and
by $(2.7)$. It is also checked readily that $A^*$
belongs to $S(\Sigma,\bar k\otimes {\rm End}\hskip 1pt L)$.
In this way, the deformation ${\sans A}'$ is characterized by a pair
$(\mu,A^*)$ formed by a Beltrami differential $\mu$ and an element $A^*$ of
$S(\Sigma,\bar k\otimes {\rm End}\hskip 1pt L)$. Conversely, to any
such pair $(\mu,A^*)$ one can associate a deformation
${\sans A}'={\sans A}(\mu,A^*)$ of the holomorphic structure.
Let us show this briefly. As explained in sect. 2, $\mu$
determines a holomorphic structure ${\sans a}'=\{z'{}_{a'}\}$ of $\Sigma$ by
solving the Beltrami equation $(2.5)$ subject to the local invertibility
condition $(2.1)$. Now, consider the equation
$$\big(\bar\partial_a-\mu_a\partial_a-(A^*{}_a-\mu_aA_a)\big)V_{aa'}=0,
\eqno(5.7)$$
with $V_{aa'}$ a smooth ${\rm GL}(r,{\bf C})$--valued function. The solution
of $(5.7)$ is determined up to right multiplication by a local smooth
${\rm GL}(r,{\bf C})$--valued function $W_{aa'}$ such that $(\bar\partial_a
-\mu_a\partial_a)W_{aa'}=0$. By using this remark. it is not difficult to see
that, if the domains of all trivializations involved intersect, one has
$V_{aa'}=L_{ab}V_{bb'}W_{aba'b'}{}^{-1}$ for some local smooth ${\rm GL}(r,
{\bf C})$--valued function $W_{aba'b'}$
satisfying $(\bar\partial_a-\mu_a\partial_a)
W_{aba'b'}=0$. From here, recalling $(2.7)$, it can be verified that,
for fixed $a'$, the matrix function $W^{a'}{}_{ab}=W_{aba'a'}$ defines an
${\sans a}'$--holomorphic ${\rm GL}(r,{\bf C})$--valued $1$--cocycle $W^{a'}$
on the domain of $z'{}_{a'}$. Since the latter is an open Riemann surface,
$W^{a'}$ is holomorphically trivial
\ref{36}. It then follows that the $V_{aa'}$'s
can be chosen so that, if the domains of all trivializations involved
intersect, one has $V_{aa'}=L_{ab}V_{bb'}L'{}_{a'b'}{}^{-1}$. Recalling
$(2.7)$ once more, it is easily checked that the matrix functions
$L'{}_{a'b'}$ define an ${\sans a}'$--holomorphic ${\rm GL}(r,
{\bf C})$--valued $1$--cocycle. Introducing trivializing maps $u'{}_{a'}$
by means of $(5.5)$, it is apparent that the collection $\{(z'{}_{a'},
u'{}_{a'})\}$ constitutes a holomorphic structure of $E$ and that the
fields $\mu$ and $A^*$ associated to it via $(2.3)$ and $(5.6)$
are precisely the ones which one started with. In conclusion,
it has been shown that the  pairs $(\mu,A^*)$ formed by a Beltrami
differential $\mu$ of ${\rm Beltr}(\Sigma)$ and an element $A^*$ of
$S(\Sigma,\bar k\otimes {\rm End}\hskip 1pt L)$
parametrize in one-to-one fashion the family of deformations
${\sans A}'$ of the holomorphic structure of $E$,
with ${\sans A}'={\sans A}$ for $\mu=0$ and $A^*=0$.
Such parametrization, however, is not unique depending on the choice of the
reference connection $A$ entering in the definition of $A^*$. Note that
$S(\Sigma,\bar k\otimes {\rm End}\hskip 1pt L)$, like
${\rm Beltr}(\Sigma)$, is an infinite dimensional holomorphic manifold.
$V$ is non local holomorphic functional of $\mu$ and $A^*$.

For a given holomorphic structure $\sans A$, the  relevant field
spaces are of the form $S(\Sigma$,
$k^{\otimes m}\otimes\bar k{\vphantom k}^{\otimes\bar m}
\otimes L^{\otimes P}\otimes\bar L^{\otimes\bar P}\otimes
L^{\vee\otimes Q}\otimes\bar L^{\vee\otimes\bar Q})$
where $m,\bar m$ $\in {\bf Z}/2$ and $P,\bar P,Q,\bar Q\in
{\bf N}\cup\{0\}$. Fields of such type provide a natural
generalization of ordinary conformal fields in extended conformal
field theory.

The intertwiners $\lambda$ and $V$ associated to a deformation
${\sans A}'$ of the holomorphic structure of $E$
induce a linear isomorphism of
$S(\Sigma,k^{\otimes m}\otimes\bar k{\vphantom k}^{\otimes\bar m}
\otimes L^{\otimes P}\otimes\bar L^{\otimes\bar P}\otimes
L^{\vee\otimes Q}\otimes\bar L^{\vee\otimes\bar Q})$
and $S(\Sigma,k'^{\otimes m}\otimes\bar k'{\vphantom k}^{\otimes\bar m}
\otimes L'^{\otimes P}\otimes\bar L'^{\otimes\bar P}
\otimes L'^{\vee\otimes Q}\otimes\bar L'^{\vee\otimes\bar Q})$,
explicitly given by
$$\Psi'{}_{a'}=(V^{\otimes-P}\otimes \bar V^{\otimes-\bar P})_{aa'}
\lambda^{\otimes-m}{}_{aa'}\bar\lambda^{\otimes-\bar m}{}_{aa'}\Psi_a
(V^{\otimes Q}\otimes\bar V^{\otimes\bar Q})_{aa'}.
\eqno(5.8)$$
This relation generalizes $(2.9)$.

The notion of Hermitian metric on a surface has a suitable generalization
to the category of complex vector bundles.
A Hermitian structure on $E$ subordinated to the holomorphic structure
$\sans A$ is a pair $(h^\circ, H)$, where $h^\circ$ is a Hermitian
metric of $\Sigma$ subordinated to $\sans a$ and $H$ is a Hermitian
metric of $L$, that is an element of $S(\Sigma,L\otimes\bar L)$ such that
$H_a=H_a{}^\dagger>0$ for any trivialization of $\sans A$.

Upon choosing a reference Hermitian structure $(g^\circ,G^\circ)$
subordinated to $\sans A$, one can express any other Hermitian
structure $(h^\circ, H)$ in terms of a Liouville field $\phi^\circ$
and a further field $\Phi$, called the Donaldson field.
The metric $h^\circ$ is given by $(2.10)$ as a functional of
$\phi^\circ$ and $g^\circ$. Similarly, $H$ can be represented
locally in the form
$$H_a=\exp\Phi_a G_a.\eqno(5.9)$$
$\Phi=\Phi(h^\circ,H)$ is an element of $S(\Sigma,{\rm End}\hskip1pt L)$
satisfying the $G$--Hermiticity condition
$$G_a\Phi_a{}^\dagger G_a{}^{-1}=\Phi_a.\eqno(5.10)$$
This ensures the Hermiticity of $H$. The pair $(\phi^\circ,\Phi)$
parametrize the family of Hermitian structures $(h^\circ,H)$ viewed as
deformations of the reference Hermitian structure $(g^\circ,G)$.

To any Hermitian structure $(h^\circ,H)$, there is canonically associated
the $(1,0)$ affine connection $\gamma_{h^\circ}$ given by $(2.11)$ and
a $(1,0)$ connection $\Gamma_H$ of $L$ compatible with $H$ given locally
by
$$\Gamma_{Ha}=\partial_aH_aH_a{}^{-1}.\eqno(5.11)$$
The curvature of $\gamma_{h^\circ}$ is given by $(2.12)$ while
the curvature of $\Gamma_{Ha}$ is
$$F_{Ha}=\bar\partial_a\Gamma_{Ha}.\eqno(5.12)$$
$F_H$ belongs to $S(\Sigma,k\otimes\bar k\otimes{\rm End} L)$.
I shall denote by $\partial_{h^\circ}$ and $\partial_H$ the
covariant derivatives associated respectively to $\gamma_{h^\circ}$
and $\Gamma_{Ha}$.

Any deformation ${\sans A}'$ of the holomorphic structure $\sans A$ of $E$
induces a deformation $(h^{\circ\prime},H')$ of a Hermitian structure
$(h^\circ,H)$, via $(2.9)$ and $(5.8)$: $h^{\circ\prime}{}_{a'}
=h^\circ{}_a|\lambda_{aa'}|^{-2}$ and $H'{}_{a'}
=V_{aa'}{}^{-1}H_aV_{aa'}{}^{-1\dagger}$.
The Liouville field $\phi^\circ$ is invariant under deformations,
the Donaldson field is not: $\Phi'{}_{a'}=V_{aa'}{}^{-1}\Phi_aV_{aa'}$.
\vskip.4cm
$c)$ {\it The relation between the holomorphic and Hermitian geometry of
E and those of its determinant and projectivization.}
\vskip.4cm
To the vector bundle $E$, there are canonically associated its determinant
line bundle $\det E$ and its projectivization $\proj E$. Such bundles
will play a role in the geometrical interpretation of the field
theoretic constructions of sects. 7, 8 and 9. In fact,
the anomalies and the relevant counterterms of extended
conformal field theory consistently appear as sums of two terms
corresponding to each of them. The determinant part of the anomaly
contributes to the ordinary conformal anomaly. The projective part gives rise
to the anomalies of extended gravity. The relevance of projective geometry
of vector bundles in extended gravity is evident in the formulations
of light cone $W$ geometry based on the parametrization of the generalized
projective structures of a surface $\Sigma$ \ref{18-20}. The changes
of trivialization of a projective bundle are indeed expressible
by means of linear fractional transformations as for projective
structures.

It is therefore important to understand the relation between the
holomorphic and Hermitian geometry of $E$ and those of $\det E$ and
$\proj E$. The analysis is based on the observation
that the covariant functors det and proj map any structure defined in $E$
into an analogous structures of $\det E$ and $\proj E$.

To each holomorphic structure $\sans A$ of $E$, one can associate
canonically a holomorphic structure of $\det E$ which will be denoted
by $\det{\sans A}$. The generic trivialization of $\det{\sans A}$ is
the image through the functor det of some trivialization $(z_a,u_a)$ of
$\sans A$ and explicitly reads as $(z_a,\det u_a)$, where $\det u_a$ is the
ordinary determinant of $u_a$ viewed as a linear isomorphism from each
fiber onto ${\bf C}^r$. The holomorphic structure of $\Sigma$ underlying
$\det{\sans A}$ is precisely $\sans a$. The associated $\sans a$--holomorphic
${\bf C}^*$ $1$--cocycle is $\det L=\{\det L_{ab}\}$.
When $r>1$, the map ${\sans A}\rightarrow\det{\sans A}$ is many-to-one.
Two holomorphic structures $\sans A$ and ${\sans A}'$ of $E$ induce the
same holomorphic structure of $\det E$ if and only if there is a
bijection $a\rightarrow a'=\varsigma(a)$ and a collection $\{Y_a\}$ of
${\rm SL}(r,{\bf C})$--valued local functions such that
$z_a=z'{}_{a'}$ and $u'{}_{a'}=Y_a\circ\pi u_a$.

Any deformation ${\sans A}'$ of the holomorphic structure
of $E$ induces a deformation $\det{\sans A}'$ of the corresponding
holomorphic structure of $\det E$. The intertwiners associated to the
latter are $\lambda$ and $\det V=\{\det V_{aa'}\}$. For any two
deformations ${\sans A}'$ and ${\sans A}''$, $\det{\sans A}'
=\det{\sans A}''$ if and only if $\mu({\sans a}')=\mu({\sans a}'')$
and $\tr A^*({\sans A}')=\tr A^*({\sans A}'')$. This follows easily
from the remarks at the end of the previous paragraph and from $(5.5)$
and $(5.6)$. It is easy to show from here that the pairs $(\mu,a^*)$ formed
by a Beltrami differential $\mu$ and an element $a^*$ of $S(\Sigma,\bar k)$
parametrize in one-to-one fashion the family of induced deformations
$\det{\sans A}'$. $a^*$ is precisely the trace
of the element $A^*$ of $S(\Sigma,\bar k\otimes {\rm End}\hskip 1pt L)$
corresponding any chosen holomorphic structure ${\sans A}'$ inducing
$\det{\sans A}'$ with a conventional $1/r$ factor: $a^*{}_b
={1\over r}\tr A^*{}_b$.

A parallel treatment can be made for the projective bundle $\proj E$.
To each holomorphic structure $\sans A$ of $E$
there is canonically associated also a holomorphic
structure of $\proj E$, which will be denoted by $\proj{\sans A}$.
The notion of holomorphic structure of a complex projective bundle over
$\Sigma$ is defined analogously to that of holomorphic structure of a
complex vector bundle but the associated holomorphic $1$--cocycle is
valued in the group of projective isomorphisms of ${\bf C}P^{r-1}$.
The generic trivialization of $\proj{\sans A}$ is the image through the
functor proj of some trivialization $(z_a,u_a)$ of $\sans A$ and
explicitly reads $(z_a,\proj u_a)$, where $\proj u_a$ is the
projective isomorphism associated canonically to $u_a$ viewed as a linear
isomorphism from each fiber onto ${\bf C}^r$. As for $\det E$,
the holomorphic structure of $\Sigma$ underlying $\proj {\sans A}$ is
precisely $\sans a$. The associated $\sans a$--holomorphic
${\bf C}^*$ $1$--cocycle is $\proj  L=\{\proj  L_{ab}\}$.
The map ${\sans A}\rightarrow\proj{\sans A}$
is many-to-one. Two holomorphic structures $\sans A$ and ${\sans A}'$ of $E$
induce the same holomorphic structure of $\proj E$ if and only if there is a
bijection $a\rightarrow a'=\varsigma(a)$ and a collection $\{y_a\}$ of
${\bf C}^*$--valued local functions such that $z_a=z'{}_{a'}$
and $u'{}_{a'}=y_a\circ\pi u_a$.
and deformations $\det{\sans A}'$ and $\proj{\sans A}'$
of $\det{\sans A}$ and $\proj {\sans A}$, respectively.

A deformation ${\sans A}'$ of the holomorphic structure of $E$ induces
a deformation $\proj{\sans A}'$ of the corresponding holomorphic structure of
$\proj E$. The intertwiners associated to the latter are $\lambda$ and
$\proj V=\{\proj V_{aa'}\}$. For any two deformations ${\sans A}'$ and
${\sans A}''$, $\proj{\sans A}'=\proj{\sans A}''$ if and only if
$\mu({\sans a}')=\mu({\sans a}'')$ and $A^*({\sans A}')-(1/r)\tr A^*(
{\sans A}')1=A^*({\sans A}'')-(1/r)\tr A^*({\sans A}'')1$. This
follows easily from the remarks at the end of the previous paragraph
and from $(5.5)$ and $(5.6)$.
It is easy to show from here that the pairs $(\mu,\hat A^*)$ formed
by a Beltrami differential $\mu$ and a traceless element $\hat A^*$ of
$S(\Sigma,\bar k\otimes{\rm End}\hskip 1pt L)$ parametrize in one-to-one
fashion the family of induced deformations. $\hat A^*$ is precisely the
traceless part of the section $A^*$ of $S(\Sigma,\bar k\otimes{\rm End}
\hskip 1pt L)$ pertaining any chosen holomorphic structure ${\sans A}'$
inducing $\proj{\sans A}'$: $\hat A^*{}_b=A^*{}_b-(1/r)\tr A^*{}_b 1$.

{}From the above, it appears that the deformations
$\det{\sans A}'$ and $\proj{\sans A}'$ induced by a deformation
${\sans A}'$ of the holomorphic structure of $E$ completely
characterize it, since, from the knowledge of the former two,
it is possible to uniquely reconstruct the latter.

The results of the last five paragraphs have a straightforward
extension to the Hermitian geometry of $\det E$ and $\proj E$,
as now I shall show.

To any Hermitian structure $(h^\circ,H)$ subordinated to the holomorphic
structure $\sans A$, there is associated a Hermitian structure $(h^\circ,
\det H)$ of $\det E$ subordinated to the induced holomorphic structure
$\det{\sans A}$. The map $(h^\circ,H)\rightarrow(h^\circ,\det H)$
is clearly many-to-one. Two Hermitian structures $(h^\circ{}_1,H_1)$
and $(h^\circ{}_2,H_2)$ of $E$ induce the same Hermitian structure  on
$\det E$ if and only if $\phi^\circ(h^\circ{}_1,H_1)
=\phi^\circ(h^\circ{}_2,H_2)$
and $\tr\Phi(h^\circ{}_1,H_1)=\tr\Phi(h^\circ{}_2,H_2)$.
Hence, the Hermitian structures $(h^\circ,H)$ of $E$ with assigned induced
Hermitian structure $(h^\circ,\det H)$ are parametrized by the pairs
$(\phi^\circ,\phi)$ of real valued elements $S(\Sigma,1)$:
$\phi_a=(1/r)\tr\Phi_a$ for any structure $(h^\circ,H)$ inducing
$(h^\circ,\det H)$.

Similarly, to any Hermitian structure $(h^\circ,H)$ subordinated to
the holomorphic structure $\sans A$, there is associated
a Hermitian structure $(h^\circ,\proj H)$
of $\proj E$ subordinated to the induced holomorphic
structure $\proj{\sans A}$, where $\proj H$ is the projective
isomorphism associated to $H$ when the latter is viewed as an
isomorphism of the antidual $\bar E^\vee$ of $E$ onto $E$.
The map $(h^\circ,H)\rightarrow(h^\circ,\proj H)$
is also many-to-one. Two Hermitian structures $(h^\circ{}_1,H_1)$
and $(h^\circ{}_2,H_2)$ of $E$ induce the same Hermitian structure  on
$\proj E$ if and only if $\phi^\circ(h^\circ{}_1,H_1)
=\phi^\circ(h^\circ{}_2,H_2)$
and $\Phi(h^\circ{}_1,H_1)-(1/r)\tr\Phi(h^\circ{}_1,H_1)1
=\Phi(h^\circ{}_2,H_2)-(1/r)\tr\Phi(h^\circ{}_2,H_2)1$.
The Hermitian structures $(h^\circ,H)$ of $E$ with assigned induced
Hermitian structure $(h^\circ,\proj H)$ are parametrized by the pairs
$(\phi^\circ,\hat\Phi)$, where $\phi^\circ$ is a real valued element of
$S(\Sigma,1)$ and a $\hat\Phi$ is a traceless element of
$S(\Sigma,{\rm End}\hskip1pt L)$ satisfying the Hermiticity condition
$(5.10)$: $\hat\Phi_a=\Phi_a-(1/r)\tr\Phi_a1$ for any structure
$(h^\circ,H)$ inducing $(h^\circ,\proj H)$.

It follows that the Hermitian structures $(h^\circ,\det H)$ and $(h^\circ,
\proj H)$ of $\det E$ and $\proj E$ induced by a Hermitian structure
$(h^\circ,H)$ of $E$ completely characterize the latter.
\vskip.4cm
$d)$ {\it Special holomorphic and Hermitian geometry of E.}
\vskip.4cm
The vector bundles relevant in ordinary conformal field theory are of the
form $E=\epsilon^{\otimes m}$ with $m\in{\bf Z}/2$. The deformations
of the holomorphic structure of such bundles are parametrized in
one-to-one fashion by the Beltrami differential $\mu$ and by the field
$A^*$. Such geometrical setting is not suitable for conformal field theory,
since, in addition to $\mu$, it contains $A^*$, an extra undesired geometrical
field. Similarly, for a general Hermitian structure $(h^\circ,H)$ of such
bundles, there is no {\it a priori} relation between $h^\circ$ and $H$,
while in conformal field theory one assumes that $H=h^{\circ\otimes m}$.
The problem just highlighted presents itself
also for more general vector bundles $E$ in extended conformal field
theory, as will be shown in sect. 7. From these simple remarks, it appears
that, in field theoretic applications, it is necessary to impose
restrictions on the class of allowed deformations and Hermitian
structures. Such restrictions are described in this subsection.

For a generic holomorphic structure $\sans A$,
the topological relation $(5.1)$ may be written as
$$\det L_{ab}{}^p=k^{\otimes q}{}_{ab}e_{ab},\eqno(5.13)$$
where $e$ is a degree zero $\sans a$--holomorphic ${\bf C}^*$--valued
$1$--cocycle. The appearance of the cocycle $e$ is unavoidable as $e$,
albeit smoothly trivial, is in general not holomorphically so. However,
as well-known, any such $1$--cocycle $e$ is holomorphically equivalent
to a flat $U(1)$--valued cocycle, i. e. one whose elements are constant
phases \ref{36}. From here, it is easy to see that any holomorphic structure
$\sans A$ of $E$ admits a reduction ${\sans A}_0$ for which
the $\sans a$--holomorphic function $e_{ab}$ appearing in $(5.13)$
is a constant of unit modulus.

Consider a deformation ${\sans A}'$ of a reference
holomorphic structure $\sans A$.
Using the analog of $(5.13)$ for ${\sans A}'$ and the relations
$\lambda_{aa'}=k_{ab}\lambda_{bb'}k'{}_{a'b'}{}^{-1}$ and $V_{aa'}=
L_{ab}V_{bb'} L'{}_{a'b'}{}^{-1}$, it is straightforward to show
that
$$\det V_{aa'}{}^p=\lambda^{\otimes q}{}_{aa'}\exp(-pr\sigma_{aa'}),
\eqno(5.14)$$
where $\exp(-pr\sigma)$ is a smooth section of the smooth ${\bf C}^*$--valued
$1$--cocycle $e\otimes e'^{-1}$. $\sigma_{aa'}$ is defined up to integer
multiples of $2\pi i/pr$. I have extracted a factor $pr$ from $\sigma_{aa'}$
in order to render its definition independent from the conventional
choice of $p$ and $q$ made above.
$\sigma$ is simply related to the normalized trace $a^*$ of $A^*$
$$a^*{}_b=\big(j\partial_b+a_b\big)\mu_b
-(\bar\partial_b-\mu_b\partial_b)\sigma_{bb'},\quad
a_b={1\over r}\tr A_b,\eqno(5.15)$$
as follows directly from $(5.6)$ and $(5.14)$.

It is possible to choose
the reference holomorphic structure $\sans A$ so that
$e_{ab}=1$ for the trivializations of the reduction ${\sans A}_0$, for the
$1$--cocycle $e$ is smoothly trivial.
This conventional choice will be assumed below.
A deformation ${\sans A}'$ is said special if
$$\sigma_{aa'}=0\quad {\rm modulo}~2\pi i/pr{\bf Z},
\quad ({\rm special~deformations})\eqno(5.16)$$
for the trivializations of the reductions ${\sans A}_0$ and ${\sans A}'_0$.
In that case, $e'{}_{a'b'}=1$, for the trivializations of ${\sans A}'_0$.
By $(5.15)$, a deformation ${\sans A}'$ is special if and only if
$a^*{}_b=\big(j\partial_b+a_b\big)\mu_b$. The necessity follows from
the fact that a flat $1$--cocycle has a holomorphic section if and
only if the cocycle is trivial \ref{35}. Thus, for a special
deformation, $a^*$ is a given functional of $\mu$. Equivalently,
$\det{\sans A}'$ is completely determined by ${\sans a}'$.
Thus, the class of special deformations is parametrized in one-to-one
fashion by the pairs $(\mu,\hat A^*)$ formed by a Beltrami field $\mu$
and a traceless element $\hat A^*$ of $S(\Sigma,\bar k\otimes{\rm End}
\hskip1pt L)$. This means that there is a one-to-one correspondence
between the family of special deformations ${\sans A}'$ and the family
of induced deformations $\proj{\sans A}'$. Indeed, there is precisely
one special deformations ${\sans A}'$ inducing $\proj{\sans A}'$.

For bundles $E$ of the form $E=\epsilon^{\otimes m}$ with $m\in{\bf Z}/2$,
the special deformations are parametrized by $\mu$ alone, since $\hat A^*=0$
automatically for the case considered. Thus, restricting to special
deformations cures the disease mentioned at the beginning of this subsection
and renders the geometric framework suitable for conformal field theory.
The need for special deformations also in extended conformal field theory
will appear in due corse in sect. 7.

The above discussion has a close counterpart for Hermitian structures.
It follows from $(5.13)$ that, for any Hermitian structure
$(h^\circ,H)$ subordinated to a holomorphic structure $\sans A$,
$$\det H_a{}^p=h^{\circ\otimes q}{}_a\exp(-pr\tau_a),\eqno(5.17)$$
where $\exp(-pr\tau)$ is a real valued element of
$S(\Sigma,e\otimes\bar e)$. This relation entails the
identity
$$\phi_a=j\phi^\circ{}_a-\tau_a(h^\circ,H)+\tau_a(g^\circ,G),
\eqno(5.18)$$
where the definition of the notation should be obvious.

A Hermitian structure $(h^\circ,H)$ of $E$ subordinated to a
holomorphic structure $\sans A$ is said special if
$$\tau_a=0,\quad ({\rm special~Hermitian~structures})\eqno(5.19)$$
for the trivializations of ${\sans A}_0$. Special Hermitian
structures clearly do exist. It is convenient to choose the
reference Hermitian structure $(g^\circ,G)$ to be of such type
and this choice will be assumed in what follows.
By $(5.18)$, a Hermitian structure $(h^\circ,H)$
is special if and only if $\phi_a=j\phi^\circ{}_a$. Thus, for a
special Hermitian structure, $\phi$ is a given functional of
$\phi^\circ$. Equivalently, the induced structure $(h^\circ,\det H)$ is
completely determined by $h^\circ$. So,
the family of special Hermitian structures is parametrized in
one-to-one fashion by the pairs $(\phi^\circ,\hat\Phi)$, where
$\hat\Phi$ is a traceless element of $S(\Sigma,{\rm End}\hskip1pt L)$
satisfying $(5.10)$. This means that there is a one-to-one correspondence
between the family of special Hermitian structures $(h^\circ,H)$ and the
family of induced structures $(h^\circ,\proj H)$. Further, there is
precisely one special Hermitian structures $(h^\circ,H)$ inducing
$(h^\circ,\proj H)$.

For bundles $E$ of the form $E=\epsilon^{\otimes m}$ with $m\in{\bf Z}/2$,
the special Hermitian structures $(h^\circ,H)$ of $E$ subordinated to a
holomorphic structure $\sans A$ are such that $H_a=h^{\circ\otimes m}{}_a$
for the trivializations of ${\sans A}_0$. This is precisely what is
required in conformal field theory. In sect. 7, it will be shown
that it is natural to restrict to special Hermitian structures
also for more general bundles $E$ in extended conformal field theory.

If $(h^\circ,H)$ is a special Hermitian structure of $E$ with respect
to a holomorphic structure $\sans A$, then, for an arbitrary deformation
${\sans A}'$ of $\sans A$, the deformed structure $(h^{\circ\prime},H')$
needs not be special. However, if ${\sans A}'$ is a special deformation,
then $(h^{\circ\prime},H')$ is special.
\vskip.6cm
\item{6.} {\bf The automorphism and extended Weyl groups and the Slavnov
operator.}
\vskip.4cm
\par
The next step is naturally the analysis of the underlying symmetry of the
geometrical setting described in the previous section.  As will be
shown below, the symmetry group is the automorphism group of the bundle
$E$ extended by the generalized Weyl group of $E$. This section is
devoted to its study.
The structure of the section parallels to some extent
that of the previous one. The section is divided in four subsections.
Subsect. $a$ provides basic generalities about the automorphism group of $E$
and describes the action of the automorphism and Weyl group on the
holomorphic and Hermitian structures of $E$. Subsect. $b$ analyzes
the relation between the symmetry group of $E$ and those of $\det E$
and $\proj E$. Subsect.
$c$ is concerned with the special symmetry subgroup preserving
special holomorphic and Hermitian structures of $E$. Finally,
subsect. $d$ introduces the infinitesimal formulation and the
Slavnov operator relevant in field theoretic applications.
\vskip.4cm
$a)$ {\it The automorphism and generalized Weyl groups of E and their action
on the holomorphic and Hermitian structures.}
\vskip.4cm
An automorphism $\alpha$ of $E$ is an orientation preserving
isomorphism of $E$ onto itself \ref{40-41}. Here, I shall restrict
to automorphisms homotopically connected to the identity ${\rm id}_E$
in the $C^\infty$ topology. The set ${\rm Aut}_c(E)$ of all automorphisms
of $E$ of such type
is clearly a group under composition of maps. As explained at the
beginning of sect. 5, to any automorphism $\alpha$, there is associated
a diffeomorphism $f_\alpha$ of $\Sigma$, which is also orientation
preserving. The map $\alpha\rightarrow f_\alpha$ is a group
homomorphism of ${\rm Aut}_c(E)$ onto
${\rm Diff}_c(\Sigma)$. Its kernel ${\rm Gau}_c(E)$ is the group of fiber
preserving automorphisms, usually called (small) gauge transformations.
${\rm Gau}_c(E)$ is a closed normal subgroup of ${\rm Aut}_c(E)$ and one
has the isomorphism ${\rm Diff}_c(\Sigma)\cong{\rm Aut}_c(E)/
{\rm Gau}_c(E)$.

If $\alpha$ is an automorphism of $E$, then, for any point $p$ of
$\Sigma$, $\alpha\big|_{E_p}$ is a linear isomorphism of $E_p$ onto
$E_{f_\alpha(p)}$. So, for any pair of trivialization $\{(z_a,u_a)\}$ and
$\{(z_b,u_b)\}$ of a reference holomorphic structure $\sans A$ such
that $f_\alpha(\dom z_a)\cap\dom z_b\not=\emptyset$, there exists a
smooth ${\rm GL}(r,{\bf C})$--valued function $\alpha_{ba}$ such that
$$u_b\circ\alpha=\alpha_{ba}\circ\pi u_a.\eqno(6.1)$$
Under changes of trivialization in $\sans A$, one has
$$\alpha_{dc}=L_{db}\circ f_\alpha\alpha_{ba}L_{ca}{}^{-1}.
\eqno(6.2)$$
If $\alpha$ belongs to ${\rm Gau}_c(E)$, $f_\alpha={\rm id}_\Sigma$.
Thus, $\alpha$ can be identified with an element of $S(\Sigma,{\rm End}
\hskip 1pt L)$.

The automorphism group ${\rm Aut}_c(E)$ acts on the set of deformations
${\sans A}'$ of the holomorphic structure $\sans A$.
Given an automorphism $\alpha$ of $E$, to any trivialization $(z'{}_{b'},
u'{}_{b'})$ of ${\sans A}'$ one associates another trivialization
$(z''{}_{a''},u''{}_{a''})$, where $z''{}_{a''}$ is given by $(3.1)$
with $f=f_\alpha$ and \ref{26}
$$u''{}_{a''}=u'{}_{b'}\circ\alpha.\eqno(6.3)$$
The collection ${\sans A}''$ of all trivializations
$(z''{}_{a''},u''{}_{a''})$ of the above form is easily
checked to be a holomorphic structure of $E$. ${\sans A}''$
is the pull-back of ${\sans A}'$ by $\alpha$ and may be denoted by
$\alpha^*{\sans A}'$. The holomorphic structure ${\sans a}''$ of
$\Sigma$ associated to ${\sans A}''$ is precisely the
pull-back $f_\alpha{}^*{\sans a}'$ of the holomorphic
structure ${\sans a}'$ associated to ${\sans A}'$, in
the sense defined in sect. 3. The ${\sans a}''$--holomorphic
${\rm GL}(r,{\bf C})$--valued $1$--cocycle $L''=\alpha^*L'$
associated to ${\sans A}''$ is related to $L'$ by
$$(\alpha^*L')_{a''b''}=L'{}_{c'd'}\circ f_\alpha,\eqno(6.4)$$
$(z''{}_{a''},u''{}_{a''})$ and $(z''{}_{b''},u''{}_{b''})$ being the
images of $(z'{}_{c'},u'{}_{c'})$ and $(z'{}_{d'},u'{}_{d'})$ under
the pull-back action of $\alpha$, respectively.

The pull-back action of ${\rm Aut}_c(E)$ on the deformations ${\sans A}'$
induces a right action of ${\rm Aut}_c(E)$ on the intertwiner fields
$\lambda$ and $V$ (cf. sect. 5). For any automorphism $\alpha\in
{\rm Aut}_c(E)$, this is defined by
$\alpha^*\lambda({\sans A}')=\lambda(\alpha^*{\sans A}')$ and
$\alpha^*V({\sans A}')=V(\alpha^*{\sans A}')$. $\alpha^*\lambda$ is
given by $(3.2)$ with $f$ substituted by $f_\alpha$. From $(5.5)$
and the identity $\pi\circ\alpha=f_\alpha\circ\pi$, one obtains
$$(\alpha^*V)_{aa''}=\alpha_{ba}{}^{-1}V_{bb'}\circ f_\alpha.\eqno(6.5)$$
In similar fashion, a right action of ${\rm Aut}_c(E)$ is induced
on the pairs $(\mu,A^*)$ formed by a Beltrami differential $\mu$ and a
field $A^*$ of $S(\Sigma,\bar k\otimes {\rm End}\hskip 1pt L)$
parametrizing the deformations of the holomorphic structure of $E$
(cf. sect. 5),
by the relations $\alpha^*\mu({\sans A}')=\mu(\alpha^*{\sans A}')$
and $\alpha^*A^*({\sans A}')=A^*(\alpha^*{\sans A}')$. $\alpha^*\mu$
is given by $(3.3)$ with $f$ substituted by $f_\alpha$. $\alpha^*A^*$
can be computed from $(3.3)$, $(5.6)$ and $(6.5)$. The local expression
of the result is.
$$\eqalignno{
&(\alpha^*A^*)_a={1\over \partial_af_{\alpha b}+
\mu_b\circ f_\alpha\partial_a\bar f_{\alpha b}}\bigg[\big
(\partial_af_{\alpha b}\bar\partial_a\bar f_{\alpha b}
-\bar\partial_af_{\alpha b}\partial_a\bar f_{\alpha b}\big)
\alpha_{ba}{}^{-1}(A^*{}_b-\mu_bA_b)\circ f_\alpha\alpha_{ba}&\cr
&+\big(\bar\partial_af_{\alpha b}+
\mu_b\circ f_\alpha\bar\partial_a\bar f_{\alpha b}\big)
\big(\alpha_{ba}{}^{-1}\partial_a\alpha_{ba}+A_a\big)
-\big(\partial_af_{\alpha b}+
 \mu_b\circ f_\alpha\partial_a\bar f_{\alpha b}\big)
\alpha_{ba}{}^{-1}\bar\partial_a\alpha_{ba}\bigg].&\cr
&&(6.6)\cr}$$

${\rm Aut}_c(E)$ acts on the space
$S(\Sigma,k^{\otimes m}\otimes\bar k{\vphantom k}^{\otimes\bar m}\otimes
 L^{\otimes P}\otimes\bar L^{\otimes\bar P}
\otimes L^{\vee\otimes Q}\otimes\bar  L^{\vee\otimes\bar Q})$.
The action is such that
$$(\alpha^*\Psi)''{}_{a''}=\Psi'{}_{b'}\circ f_\alpha,\eqno(6.7)$$
where ${\sans A}''=\alpha^*{\sans A}'$. From $(2.8)$, $(3.6)$,
$(5.8)$ and $(6.5)$, one finds
$$\alpha^*\Psi_a=(\alpha^{\otimes-P}\otimes\bar\alpha^{\otimes-\bar P})_{ba}
\varpi_{ab}(f_\alpha;\mu)^{2m}\bar\varpi_{ab}(f_\alpha;\mu)^{2\bar m}
\Psi_b\circ f_\alpha(\alpha^{\otimes Q}\otimes\bar\alpha^{\otimes
\bar Q})_{ba}.\eqno(6.8)$$

When Hermitian structures on $E$ are envisaged, the symmetry group must be
enlarged to the semidirect product ${\rm Aut}_c(E)\times{\rm Weyl}(E)$.
${\rm Weyl}(E)$ is the generalized Weyl group of $E$. ${\rm Weyl}(E)$
is the direct product group $\exp S(\Sigma,1)\times
\exp S(\Sigma,{\rm End}\hskip1pt L)$ and contains the ordinary
Weyl group ${\rm Weyl}(\Sigma)$ as a subgroup.
The group multiplication is defined by
$$\big(\alpha_1,\upsilon^\circ{}_1,\Upsilon_1\big)
\big(\alpha_2,\upsilon^\circ{}_2,\Upsilon_2\big)=
\big(\alpha_1\circ \alpha_2,\upsilon^\circ{}_1
f_{\alpha_1}{}^{-1*}\upsilon^\circ{}_2,\Upsilon_1
\alpha_1{}^{-1*}\Upsilon_2\big).\eqno(6.9)$$

The action of a combined automorphism and extended Weyl transformation
$(\alpha,\upsilon^\circ$, $\Upsilon)$
on the fields $\lambda$, $V$, $\mu$ and $A^*$ and on the generalized
conformal fields reduces to that of $\alpha$ given above.
On a Hermitian structure $(h^\circ,H)$, it
is given by $(3.8)$ with $h$, $f$ and $\upsilon$ replaced by
$h^\circ$, $f_\alpha$ and $\upsilon^\circ$ and by
$$\alpha^*{\vphantom)}^{\upsilon^\circ\Upsilon}H=
\alpha^*\big(\Upsilon^{-1}H\Upsilon^{-1\dagger}\big).\eqno(6.10)$$
This may be translated into an action on the Donaldson field $\Phi$,
though I have found no simple explicit expression.
\vskip.4cm
$b)$ {\it The relation between the symmetry group of E and those of its
determinant and projectivization.}
\vskip.4cm
To each automorphism $\alpha$ of $E$, there is associated an
automorphism $\det\alpha$ of $\det E$ by the functor det. From the
covariance of det, relations analogous to $(6.1)$--$(6.2)$ are found
to hold for $\det u_a$ and $\det\alpha_{ba}$. The map $\alpha\rightarrow
\det\alpha$ is a many-to-one homomorphism of ${\rm Aut}_c(E)$ onto
${\rm Aut}_c(\det E)$. Two automorphisms $\alpha$ and
$\beta$ of $E$ induce the same automorphism of $\det E$ if and only if
$f_\alpha=f_\beta$ and $\det\alpha_{ba}=\det\beta_{ba}$ for all $a,b$'s.
Similarly, to each automorphism $\alpha$ of $E$, there is associated
an automorphism $\proj\alpha$ of $\proj E$ by the functor proj
with completely analogous properties. It is not difficult to show
that the automorphisms $\det\alpha$ and $\proj\alpha$ completely
characterize $\alpha$. The proof of this property uses
crucially the fact that the automorphisms considered are
homotopically connected to the identity.

For any automorphism $\alpha$ and any deformation ${\sans A}'$ of
the holomorphic structure of $E$, the pull-back by the automorphism
$\det\alpha$ of the induced deformations $\det{\sans A}'$ of $\det E$
is such that $(\det\alpha)^*\det{\sans A}'=\det(\alpha^*{\sans A}')$.
By the covariance of the functor det, the trivializations of
$(\det\alpha)^*\det{\sans A}'$ and its associated cocycle
$\det\alpha^*\det L'$ obey relations analogous to $(6.3)$--$(6.4)$.
Further, $(6.5)$ implies an analogous relation for the deformation
intertwiners $\det V$. The action of $\det\alpha$ on
the induced deformations $\det{\sans A}'$ translates into an
action on the fields $(\mu,a^*)$ parametrizing the latter (cf. sect. 5).
In particular, $\det\alpha^*a^*$ is given by $(6.6)$ with $A^*$,
$A$ and $\alpha_{ba}$ replaced by $a^*{}_b$, $a_b=(1/r)\tr A_b$
and $\det\alpha_{ba}$, respectively.
Totally analogous properties hold for the pull-back action of $\proj
\alpha$ on the induced deformations $\proj{\sans A}'$.
In particular, $\proj\alpha^*\hat A^*$ is given by $(6.6)$
with $A^*{}_a$, $A_a$ and $\alpha_{ba}$ replaced by $\hat A^*{}_a$,
$\hat A_a=A_a-(1/r)\tr A_a1$ and
$(\det\alpha_{ba})^{-{1\over r}}\alpha_{ba}$, respectively.

One may similarly define the extended symmetry groups
${\rm Aut}_c(\det E)\times{\rm Weyl}(\det E)$
and ${\rm Aut}_c(\proj E)\times{\rm Weyl}(\proj E)$
of $\det E$ and $\proj E$.
The covariant functors det and proj yield group homomorphisms
from the extended symmetry group of $E$ onto those of $\det E$ and
$\proj E$, namely $(\alpha,\upsilon^\circ,\Upsilon)\rightarrow
(\det\alpha,\upsilon^\circ,\det\Upsilon)$ and
$(\alpha,\upsilon^\circ,\Upsilon)\rightarrow(\proj\alpha,\upsilon^\circ,
\proj\Upsilon)$, respectively. These homomorphisms are many-to-one.
However, $(\det\alpha,\upsilon^\circ,\det\Upsilon)$ and $(\proj\alpha$,
$\upsilon^\circ,\proj\Upsilon)$ completely characterize $(\alpha,
\upsilon^\circ,\Upsilon)$. $(\det\alpha,\upsilon^\circ,\det\Upsilon)$ and
$(\proj\alpha,\upsilon^\circ$, $\proj\Upsilon)$ act on $(h^\circ,\det H)$
and $(h^\circ,\proj H)$, respectively, and thus on $\phi$ and $\hat\Phi$.
\vskip.4cm
$c)$ {\it The special symmetry group of E.}
\vskip.4cm
In sect. 5, I have introduced the notion of special deformation and
special Hermitian structure. It is now necessary to find out which
subgroup of the symmetry group preserves the special character of
such structures. This is the topic of this subsection.

It has been shown in sect. 5 that there is a
${\bf C}^*$--valued $1$--cocycle
$e'$ associated to any deformation ${\sans A}'$ (cf. eq. $(5.13)$) and
that $e'$ has the property of being flat unitary when restricting to
a certain reduction ${\sans A}'_0$ of ${\sans A}'$. The question arises
whether such property is preserved by the action of the automorphism
group. Denoting by $\alpha^* e'$ the
$f_\alpha^*{\sans a}'$--holomorphic ${\bf C}^*$--valued $1$ cocycle
corresponding to $\alpha^* L'$, it is not
difficult to show from $(5.13)$ and $(6.4)$ that
$$(\alpha^*e)'{}_{a''b''}=e'{}_{c'd'}\circ f_\alpha,\eqno(6.11)$$
where ${\sans A}''=\alpha^*{\sans A}'$.
This relation shows that the reduction ${\sans A}'_0$ of
${\sans A}'$ with respect to whose trivializations $e'{}_{a'b'}$ is a
constant of unit modulus is mapped by $\alpha^*$ into its counterpart
${\sans A}''_0$ in ${\sans A}''$. This property is crucial for
the consistency of what follows.

{}From $(5.14)$, $(6.5)$ and $(3.2)$, one finds
$$\exp\big(pr\alpha^*\sigma_{aa''}-pr\sigma_{bb'}\circ f_\alpha\big)
=\varpi_{ab}(f_\alpha;\mu)^{2q}\det\alpha_{ba}{}^p,\eqno(6.12)$$
where $\alpha^*\sigma$ be the intertwiner field associated to
$\alpha^*V$ via $(5.14)$. From this relation and $(5.16)$,
it appears that a general automorphism
$\alpha$ does not map the subclass of special deformations into itself,
since $\alpha^*\sigma_{aa'}$ may not vanish for any reduction
of ${\sans A}''$. From $(6.12)$, the automorphisms $\alpha$ for which
this happens are those satisfying the relation
$$\det\alpha_{ba}{}^{-p}=\varpi_{ab}(f_\alpha;\mu)^{2q},\quad
({\rm special~automorphisms})\eqno(6.13)$$
for the trivializations of ${\sans A}_0$.
I shall call such automorphisms special. They form a subgroup
${\rm Aut}(E;\mu)$ of ${\rm Aut}(E)$ depending on $\mu$.
For any special automorphism $\alpha$, $\det\alpha$ is completely
determined by $f_\alpha$ and $\mu$. Since $\det\alpha$ and $\proj\alpha$
characterize $\alpha$, the restriction of the group homomorphism
$\alpha\rightarrow\proj\alpha$ to the subgroup of special automorphisms
of $E$ is a group isomorphisms.

Proceeding in similar fashion for Hermitian structures, one finds
the relation
$$\exp\big(pr\alpha^*{\vphantom\tau}^{\upsilon^\circ\Upsilon}\tau_a
-pr\tau_b\circ f_\alpha\big)=\big|\varpi_{ab}(f_\alpha;\mu)^{2q}
\det\alpha_{ba}{}^p(\upsilon^\circ{}_b{}^{-q} \det\Upsilon_b{}^p)\circ
f_\alpha\big|^2,\eqno(6.14)$$
from $(5.17)$, $(3.8)$ and $(6.10)$. From this relation and $(5.19)$,
one sees that the action of the symmetry group on the Hermitian structures
of $E$ does not map the subclass of special Hermitian structures into itself.
As appears from $(6.14)$, in order a combined automorphism and extended Weyl
transformation $(\alpha,\upsilon^\circ,\Upsilon)$ to have such a property,
it is sufficient that $\alpha$ is special and that
$$\det\Upsilon_a{}^p=\upsilon^{\circ\otimes q}{}_a,\quad
({\rm special~Weyl~transformations})\eqno(6.15)$$
for the trivializations of ${\sans A}_0$. The transformations
of such type form a special subgroup ${\rm Aut}_c(E;\mu)\times{\rm Weyl}_0(E)$
of the full extended symmetry group ${\rm Aut}_c(E)\times{\rm Weyl}(E)$,
depending on $\mu$. For a special combined automorphism and Weyl
transformation $(\alpha,\upsilon^\circ,\Upsilon)$,
$(\det\alpha,\upsilon^\circ,\det\Upsilon)$ is completely
determined by $f_\alpha$, $\mu$ and $\upsilon^\circ$. Since
$(\det\alpha,\upsilon^\circ,\det\Upsilon)$ and $(\proj\alpha,\upsilon^\circ,
\proj\Upsilon)$ completely characterize $(\alpha,\upsilon,\Upsilon)$,
the restriction of the group automorphism $(\alpha,\upsilon^\circ,\Upsilon)
\rightarrow(\proj\alpha,\upsilon^\circ,\proj\Upsilon)$
to the subgroup of special symmetry transformations is
a group isomorphism. Note how the results of the previous paragraph
generalize directly to Hermitian geometry.
\vskip.4cm
$d)$ {\it Infinitesimal formulation and the Slavnov operator.}
\vskip.4cm
As for the symmetry of surfaces, one is mainly interested in the
infinitesimal action of ${\rm Aut}_c(E)\times{\rm Weyl}(E)$
and is thus lead to considering its Lie algebra
${\rm Lie}\hskip1pt\big({\rm Aut}_c(E)\times{\rm Weyl}(E)
\big)$ \ref{26}. Writing formally $f_\alpha={\rm id}_\Sigma+\xi$,
$\alpha={\rm id}_E+M$, $\upsilon^\circ=1-\theta^\circ/2$
and $\Upsilon=1-\Theta/2$ in $(6.9)$,
one finds that ${\rm Lie}\hskip1pt\big({\rm Aut}_c(E)
\times{\rm Weyl}(E)\big)$ is isomorphic to the semidirect sum
$\big(\bigcup_{\xi\in S(\Sigma,k^{-1})}\{\xi\}\times(\xi A+S(\Sigma,{\rm End}
\hskip1pt L))\big)\oplus S(\Sigma,1)\oplus S(\Sigma,{\rm End}\hskip1pt L)$,
where $A$ is a reference $(1,0)$ connection of $L$, with Lie brackets
$$\eqalignno{
&[(\xi_1,\Omega_1)\oplus\theta^\circ{}_1\oplus\Theta_1,
(\xi_2,\Omega_2)\oplus\theta^\circ{}_2\oplus\Theta_2]=
\big((\xi_1\partial+\bar\xi_1\bar\partial)\xi_2
-(\xi_2\partial+\bar\xi_2\bar\partial)\xi_1,
&\cr&(\xi_1\partial+\bar\xi_1\bar\partial)\Omega_2
-(\xi_2\partial+\bar\xi_2\bar\partial)\Omega_1-[\Omega_1,\Omega_2]\big)
\oplus\big((\xi_1\partial+\bar\xi_1\bar\partial)\theta^\circ{}_2
-(\xi_2\partial+\bar\xi_2\bar\partial)\theta^\circ{}_1\big)&\cr&
\oplus\big((\xi_1\partial+\bar\xi_1\bar\partial)\Theta_2-(\xi_2\partial+
\bar\xi_2\bar\partial)\Theta_1-[\Omega_1,\Theta_2]+[\Omega_2,\Theta_1]
+(1/2)[\Theta_1,\Theta_2]\big).&(6.16)\cr}$$

To express the infinitesimal action of the symmetry group
${\rm Aut}_c(E)\times{\rm Weyl}(E)$ on the relevant space $\cal F$ of
field functionals, one introduces the exterior algebra
$\bigwedge^*{\rm Lie}\hskip1pt\big({\rm Aut}_c(E)\times{\rm Weyl}(E)\big)
{\vphantom )}^\vee$ of ${\rm Lie}\hskip1pt\big({\rm Aut}_c(E)\times
{\rm Weyl}(E)\big)$. The standard generators of the algebra are the
automorphism ghosts $c$ and $M$ and the Weyl ghost $w^\circ$ and $W$
\footnote{${}^7$}{In ref. \ref{26}, the ghost $M$ is written as a sum
$M_{\rm pt}+G$, where $M_{\rm pt}$ corresponds to parallel transport
along the vector field $c\partial+\bar c\bar\partial$ and
$G$ is the gauge ghost. Though interesting for other reasons, this
separation is not relevant in the following analysis.}.
$c$ is a section of $k^{-1}\otimes
\bigwedge^1{\rm Lie}\hskip1pt\big({\rm Aut}_c(E)\times
{\rm Weyl}(E)\big){\vphantom )}^\vee$, $w^\circ$ is a section of
$1\otimes \bigwedge^1{\rm Lie}\hskip1pt\big({\rm Aut}_c(E)\times
{\rm Weyl}(E)\big){\vphantom )}^\vee$, $M-cA$ and $W$ are
sections of ${\rm End}\hskip1pt L\otimes\bigwedge^1{\rm Lie}
\hskip1pt\big({\rm Aut}_c(E)\times{\rm Weyl}(E)\big)
{\vphantom )}^\vee$. $c$, $M$ $w^\circ$ and $W$ are defined by
$\langle(c(p),M(p)),(\xi,\Omega)\oplus\theta^\circ\oplus\Theta\rangle
=(\xi(p),\Omega(p))$, $\langle w^\circ(p),(\xi,\Omega)\oplus\theta^\circ
\oplus\Theta\rangle=\theta^\circ(p)$ and $\langle W(p),(\xi,\Omega)
\oplus\theta^\circ\oplus\Theta\rangle=\Theta(p)$ for $p\in\Sigma$,
respectively.

Linearizing the action of the symmetry group on the relevant space
${\cal F}$ of field functionals, defines a nilpotent coboundary
operator $s$ on the  space
$\bigwedge^*{\rm Lie}\big(\hskip1pt{\rm Aut}_c(E)\times{\rm Weyl}
(E)\big){\vphantom )}^\vee$ $\otimes{\cal F}$ called, as in sect. 2,
the Slavnov operator. From $(6.16)$, one can easily
read off the structure equations obeyed by $c$, $M$, $w^\circ$ and $W$.
For $c$ and $w^\circ$, these are $(3.10)$ and $(3.11)$, respectively.
For $M$ and $W$, they read \ref{26}
$$sM=(c\partial+\bar c\bar\partial)M-M^2,\eqno(6.17)$$
$$sW=(c\partial+\bar c\bar\partial)W-[M,W]+(1/2)W^2.\eqno(6.18)$$
For a given Beltrami differential $\mu$ in ${\rm Beltr}(\Sigma)$ and a
given element $A^*$ of $S(\Sigma,\bar k\otimes{\rm End} L)$, the relevant
combinations of ghost fields are $C$, given by $(3.12)$, and \ref{26}
$$X=M-cA-\bar cA^*,\eqno(6.19)$$
where $A$ is the same reference connection of $L$ entering the
definition of $A^*$, eq. $(5.6)$.
$X$ is a section of ${\rm End}\hskip1pt L\otimes\bigwedge^1{\rm Lie}
\hskip1pt\big({\rm Aut}_c(E)\times{\rm Weyl}(E)\big)
{\vphantom )}^\vee$.
$(3.13)$, $(3.14)$ and $(3.15)$ hold unchanged in the present more general
context. From $(6.5)$ and $(6.6)$, one has further
$$sV_{a'}=C(\partial-A)V_{a'}-XV_{a'},\eqno(6.20)$$
$$sA^*=-\big(\bar\partial-\mu\partial_A-\ad A^*\big)X+C\big(\partial A^*
-\bar\partial A+[A^*,A]\big),\quad sA=0,\eqno(6.21)$$
where $\partial_A=\partial-\ad A$ is the covariant derivative of the
connection $A$
\footnote{${}^8$}{For any two matrices $S$ and $T$ $\ad S\cdot T=[S,T]$.}.
{}From $(3.10)$, $(6.17)$ and $(6.21)$, one finds that
$$sX=C\partial_AX-X^2.\eqno(6.22)$$

{}From $(6.7)$--$(6.8)$, it follows that, for any field $\Psi$ in
$S(\Sigma,k^{\otimes m}\otimes\bar k{\vphantom k}^{\otimes\bar m}\otimes
 L^{\otimes P}\otimes\bar L^{\otimes\bar P}
\otimes L^{\vee\otimes Q}\otimes\bar  L^{\vee\otimes\bar Q})$,
one has
$$s\Psi'=(c\partial+\bar c\bar\partial)\Psi',\eqno(6.23)$$
$$\eqalignno{
s\Psi=&\hskip1pt(c\partial+\bar c\bar\partial)\Psi
+\big(m(\partial c+\mu\partial\bar c)+\bar m(\bar\partial\bar c+
\bar\mu\bar\partial c)\big)\Psi
-\big[{\cal R}_P(M)\otimes\bar 1^{\otimes\bar P}&\cr&+
1^{\otimes P}\otimes\bar{\cal R}_{\bar P}(\bar M)\big]\Psi
+\Psi\big[{\cal R}_Q(M)\otimes\bar 1^{\otimes\bar Q}+
1^{\otimes Q}\otimes\bar{\cal R}_{\bar Q}(\bar M)\big],&(6.24)\cr}$$
where ${\cal R}_R(T)=\sum_{s=0}^{R-1}1^{\otimes s}\otimes T\otimes
1^{\otimes R-s-1}$ and $\bar{\cal R}_R(\bar T)=\sum_{s=0}^{R-1}\bar
1^{\otimes s}\otimes \bar T\otimes \bar 1^{\otimes R-s-1}$. These
relations generalize $(3.16)$--$(3.17)$.

$s\phi^\circ$ is given by $(3.18)$ in terms of $\phi^\circ$ and $w^\circ$.
A simple calculation using $(5.9)$ and $(6.10)$ yields
$$\eqalignno{s\Phi=&\hskip1pt(c\partial_G+\bar c\bar\partial)\Phi+
{\ad\Phi\over\exp\ad\Phi-1}\big(c\Gamma_G-M+(1/2)W\big)&\cr
&+G\bigg({\ad\Phi\over\exp\ad\Phi-1}\big(c\Gamma_G-M+(1/2)W\big)\bigg)^\dagger
G^{-1},\quad sG=0,&(6.25)\cr}$$
where $\partial_G=\partial-\ad\Gamma_G$ and
$\ad\Phi/(\exp\ad\Phi-1)$ is understood as the power series of the
function $t\rightarrow t/(\exp t-1)$ with $t$ replaced by $\ad\Phi$.

It is simple to show that ${\rm Lie}\hskip1pt\big({\rm Aut}_c(\det E)
\times{\rm Weyl}(\det E)\big)$ and ${\rm Lie}\hskip1pt\big({\rm Aut}_c(
\proj E)\times{\rm Weyl}(\proj E)\big)$ are isomorphic to the
semidirect sums $\big(\bigcup_{\xi\in S(\Sigma,k^{-1})}\{\xi\}\times(\xi a
+S(\Sigma,1))\big)\oplus S(\Sigma,1)\oplus S(\Sigma,1)$
and $\big(\bigcup_{\xi\in S(\Sigma,k^{-1})}\{\xi\}\times(\xi\hat A+
\hat S(\Sigma,{\rm End}\hskip1pt L))\big)\oplus S(\Sigma,1)\oplus
\hat S(\Sigma,{\rm End}\hskip1pt L)$, respectively, where $a$ and
$\hat A$ are the normalized trace and traceless part of the connection
$A$ and $\hat S(\Sigma,{\rm End}\hskip1pt L)$ is the space of
traceless sections of ${\rm End}\hskip1pt L$.
The corresponding Lie brackets are given by $(6.16)$ upon
setting formally $\Omega_i=\omega_i1$ and $\Theta_i=\theta_i1$ and
$\Omega_i=\hat\Omega_i$ and $\Theta_i=\hat\Theta_i$, respectively, with
$\omega_i,\theta_i\in S(\Sigma,1)$ and $\hat\Omega_i,\hat\Theta_i\in
\hat S(\Sigma,{\rm End}\hskip1pt L)$. The covariant
functors det and proj induce the Lie algebras homomorphisms
$(\xi,\Omega)\oplus\theta^\circ\oplus\Theta\rightarrow
(\xi,\omega)\oplus\theta^\circ\oplus\theta$ and
$(\xi,\Omega)\oplus\theta^\circ\oplus\Theta\rightarrow
(\xi,\hat\Omega)\oplus\theta^\circ\oplus\hat\Theta$
of ${\rm Lie}\hskip1pt\big({\rm Aut}_c(E)\times{\rm Weyl}(E)
\big)$ onto ${\rm Lie}\hskip1pt\big({\rm Aut}_c(\det E)\times
{\rm Weyl}(\det E)\big)$ and ${\rm Lie}\hskip1pt\big({\rm Aut}_c(\proj E)
\times{\rm Weyl}(\proj E)\big)$, respectively, where $\omega=(1/r)\tr\Omega$
and $\hat\Omega=\Omega-(1/r)\tr\Omega1$ and similarly for $\Theta$.

To the Lie algebra homomorphisms defined in the previous paragraph,
there correspond a decomposition of the ghost fields $M$ and $W$ in
their normalized trace and traceless parts $m$, $\hat M$, $w$ and $\hat W$,
respectively.
$sm$, $s\hat M$, $sw$ and $s\hat W$, can then be easily found from $(6.17)$
and $(6.18)$. The relevant combinations
of ghost fields are the normalized trace $x$ and the
trace part $\hat X$ of the ghost field $X$.
$sa^*$ can be easily obtained by tracing $(6.21)$ and expressing everything
in terms of $x$, $a$ and $a^*$. $sA^*$ is given by $(6.21)$ with $X$ $A$
and $A^*$ replaced by $\hat X$ $\hat A$ and $\hat A^*$, respectively.
{}From $(6.22)$, it is easy to see that $sx=C\partial x$ and that $s\hat X$
is given by $(6.22)$ with $X$ and $A$ substituted by $\hat X$ and $\hat A$.
{}From $(6.25)$, one can also easily obtain expressions for $s\phi$ and
$s\hat\Phi$.

{}From $(6.12)$, after some simple rearrangements using $(5.15)$, one gets
$$s\sigma_{a'}=(j\partial+a+\partial\sigma_{a'})C+x.\eqno(6.26)$$
{}From $(6.14)$, one can easily find also a formula for $s\tau$.

The special symmetry transformations correspond to a subalgebra of
the symmetry algebra defined by the constraints $\omega=-j(\partial\xi
+\mu\partial\bar\xi)$ and $\theta=j\theta^\circ$. Hence, when restricting
to the special symmetry algebra, the relations
$$m=-j(\partial c+\mu\partial\bar c),\quad({\rm special~automorphisms})
\eqno(6.27)$$
$$w=jw^\circ,\quad({\rm special~Weyl~transformations})\eqno(6.28)$$
hold, as follows from $(6.13)$ and $(6.15)$.
For special deformations and restricting to the special symmetry algebra
one has that
$$x=-(j\partial+a)C, \eqno(6.29)$$
by $(5.15)$, $(5.16)$, $(6.27)$ and $(6.19)$.
\vskip.6cm
\item{7.} {\bf Extended conformal field theory: automorphism versus
extended Weyl anomaly.}
\vskip.4cm
\par
I shall now describe a generalization of conformal field theory,
trying to highlight as much as possible the parallelism with ordinary
conformal field theory.
The geometric setting is that described in sects. 5 and 6. Consider a
holomorphic structure $\sans A$ of $E$ and the associated holomorphic
structure $\sans a$ of $\Sigma$. The basic fields of a generalized conformal
field theory are smooth sections of $\sans a$--holomorphic $1$--cocycles
of the form $k^{\otimes m}\otimes L$ with $m\in{\bf Z}/2$.
The classical action is given by the integral of a $(1,1)$ density
depending locally on the fields. In this sense, one is dealing with a
classical field theory with an extended conformal symmetry.
The quantization is carried out by means of the $\zeta$ function
renormalization scheme. This requires the introduction of a Hermitian
structure $(h^\circ,H)$ of $E$ to properly define the adjoint
of the relevant differential operators. The resulting effective
action $I_{\rm aut}(h^\circ,H)$ will thus depend on this structure.
One may fix a reference holomorphic structure $\sans A$
of $E$ and the Hermitian metrics $H$ and consider any deformation
${\sans A}'$ of $\sans A$ and the deformation $(h^{\circ\prime},H')$ of
$(h^\circ,H)$ associated to ${\sans A}'$. Correspondingly, one has the
effective action $I'{}_{\rm aut}(h^{\circ\prime},H')$. In the
present context, the symmetry group is ${\rm Aut}(E)\times{\rm Weyl}(E)$.
The $\zeta$ function scheme is automorphism invariant but not Weyl invariant,
for in general $I'{}_{\rm aut}(h^{\circ\prime},H')$ depends on the metric
$H'$ and not simply on the holomorphic structure ${\sans A}'$. As for
ordinary conformal field theory, the form of the extended Weyl anomaly is
universal:
$$sI'{}_{\rm aut}(h^{\circ\prime},H')=
\kappa^\circ{\cal A}_{\rm conf}\big(w^\circ{}_1,h^{\circ\prime}\big)
+K{\cal A}'{}_{\rm ext}\big(W'_{H'},H'\big),\eqno(7.1)$$
where ${\cal A}_{\rm conf}$ and $w^\circ{}_1$ are given by $(4.2)$
and $(4.3)$ and
$${\cal A}'{}_{\rm ext}(W'_{H'},H')={-1\over 12\pi}\int_\Sigma
{d\bar z'\wedge dz'\over 2i}\tr\big(W'_{H'}F'_{H'}\big),\eqno(7.2)$$
$$W'_{H'}=\big(W'+H' W'{}^\dagger H'{}^{-1}\big)/2,\eqno(7.3)$$
$W'$ being the deformation of the Weyl ghost $W$.
The anomaly contains a contribution which is of the form of an
ordinary conformal anomaly with central charge $\kappa^\circ$.
The other contribution reflects the dependence on the scale of $H'$.
$K$ is a further coefficient measuring the anomaly strength and
may be viewed as a generalized central charge.
Note that ${\cal A}'{}_{\rm ext}$ reduces to ${\cal A}'{}_{\rm conf}$ when the
rank $r$ of $E$ equals $1$.
To check the consistency of the above, one must verify that the anomaly
${\cal A}'{}_{\rm ext}$ is a local Slavnov cohomology class.
This can be seen as follows.

The structure of the extended Weyl anomaly $(7.2)$--$(7.3)$ suggests
considering the following differential form on the space ${\cal H}'$ of
Hermitian structures of $E$ subordinated to the holomorphic structure
${\sans A}'$:
$$D'={1\over 12\pi}\int_\Sigma{d\bar z'\wedge dz'\over 2i}
\tr\big(\delta H'H'^{-1}F'_{H'}\big).\eqno(7.4)$$
Using that $\delta^2=0$, it is simple to verify that $\delta D'=0$, i. e.
$D'$ is closed.

For any field $\Psi$, define
$$s_{\rm diff}\Psi'=(c\partial+\bar c\bar\partial)\Psi',\eqno(7.5)$$
$$s_{\rm diff}c=sc=(c\partial+\bar c\bar\partial)c,\eqno(7.6)$$
(cf. eq. $(3.10)$). $s_{\rm diff}$ is a nilpotent operator,
as is easy to verify.
$s_{\rm diff}$ expresses the infinitesimal pull-back action of the
diffeomorphism group on deformations of fields. Indeed,
$s_{\rm diff}$ is the same as $s$ when acting on all those fields for which
the action of the symmetry group is just the automorphism pull-back action,
but it obviously differs from $s$ when acting on fields such as $H'$ and
$W'$ which do not enjoy such property. In fact,
recalling $(6.7)$--$(6.8)$ and $(6.10)$, it is easy to see that
$$sH'=s_{\rm diff}H'+W'_{H'}H',\eqno(7.7)$$
$$sW'=s_{\rm diff}W'+(1/2)W'^2.\eqno(7.8)$$

Since integration is invariant under the pull-back action of diffeomorphisms,
one has
$$s_{\rm diff}{\cal A}'{}_{\rm ext}(W'_{H'},H')=0.\eqno(7.9)$$
Now, from $(3.1)$, it appears that $(s-s_{\rm diff})z'=0$. So, the
computation of any expression of the form
$(s-s_{\rm diff})\int_\Sigma{d\bar z'\wedge dz'\over 2i}\tau'$
can be carried out by varying the fields but keeping the
coordinates $z'$ fixed. From this remark,
$(7.7)$ and $(7.9)$ and the closedness of $D$, one finds then
$$\eqalignno{
&s{\cal A}'{}_{\rm ext}(W'_{H'},H')=
(s-s_{\rm diff}){\cal A}'{}_{\rm ext}(W'_{H'},H')
&\cr&=(s-s_{\rm diff}){-1\over 12\pi}\int_\Sigma{d\bar z'\wedge dz'\over 2i}
\tr\big((s-s_{\rm diff})H'H'^{-1}F'_{H'}\big)=0.&(7.10)\cr}$$
{}From $(7.10)$, it follows then that ${\cal A}'{}_{\rm ext}$ is $s$--closed.
The locality of ${\cal A}'{}_{\rm ext}$ can by verified directly by expressing
it in terms of the fields $\mu$ and $A^*$ parametrizing the deformations
${\sans A}'$ and the Donaldson field $\Phi$ parametrizing the metrics $H$
with respect to a fixed reference metric $G$
(cf. sect. 5). The reference connection $(1,0)$ $A$ of $L$
entering into the definition of $A^*$ (cf. eq. $(5.6)$) is conveniently chosen
to be the metric connection $\Gamma_G$ of $G$. Using $(2.6)$, $(2.7)$,
$(5.6)$ and $(5.8)$, one finds
$$\eqalignno{&{\cal A}'{}_{\rm ext}(W'_{H'},H')=
{\cal A}_{\rm ext}(W,W^\dagger,\Phi,\mu,\bar\mu,A^*,A^{*\dagger};G)=
{-1\over 12\pi}\int_\Sigma{d\bar z\wedge dz\over 2i}\tr\bigg\{W_H
\bigg[F_H&\cr&
-\big(\partial_G-\bar\mu\bar\partial-(\bar\partial\bar\mu)\big)
{A^*{}_H\over 1-\mu\bar\mu}
-\big(\bar\partial-\mu\partial_G-(\partial\mu)\big)
{GA^*{}_H{}^\dagger G^{-1}\over 1-\mu\bar\mu}
+{[A^*{}_H,GA^*{}_H{}^\dagger G^{-1}]\over 1-\mu\bar\mu}\bigg]\bigg\},
&\cr&&(7.11a)\cr}$$
where $W_H$ is given by an expression analogous to $(7.3)$ and
$$A^*{}_H=A^*+\mu(\Gamma_H-\Gamma_G).\eqno(7.11b)$$
Recall that $H=\exp\Phi G$ (cf. eq. $(5.9)$),
$\Gamma_H=\Gamma_G+\partial_G\exp\Phi\exp-\Phi$ and
$F_H=F_G+\bar\partial(\partial_G\exp\Phi\exp-\Phi)$.

In analogy to
ordinary conformal field theory, the extended Weyl anomaly
can be absorbed by adding to $I'{}_{\rm aut}(h^{\circ\prime},H')$ a local
counterterm which in turn creates a chirally split automorphism anomaly.
I shall now expound in some detail the calculation of such counterterm.

I have shown above that the differential form $D'$ defined by $(7.4)$
is closed. Since ${\cal H}'$
is clearly contractible, $D'$ is also exact. To compute a primitive of $D'$,
one can integrate $D'$ along any functional path in ${\cal H}'$. A convenient
choice is given by the path $H'_t=\exp(t\Phi')G'$ with $0\leq t\leq 1$.
The result reads
$$S'{}_D(\Phi',G')={-1\over 12\pi}\int_\Sigma{d\bar z'\wedge dz'\over 2i}
\tr\bigg[\bar\partial'\Phi'{\exp\ad\Phi'-1-\ad\Phi'\over (\ad\Phi')^2}
\partial'_{G'}\Phi'-F'_{G'}\Phi'\bigg],\eqno(7.12)$$
where $(\exp\ad\Phi'-1-\ad\Phi')/(\ad\Phi')^2$ is the power series of the
function $(\exp t-1-t)/t^2$ with $t$ substituted by $\ad\Phi'$.
$S'{}_D(\Phi',G')$ is the Donaldson action (with no extended cosmological term)
\ref{29} and generalizes the Liouville action in the present context.
Indeed, it is easy to verify that when $E$ has rank $r=1$, $S'{}_D(\Phi',G')$
reduces to the Liouville action $S_L(\phi,g')$ (cf. eq. $(4.4)$).
It is interesting to express $S'{}_D(\Phi',G')$ directly in terms of the
fields $\mu$ $A^*$ and $\Phi$. Using $(2.6)$, $(2.7)$, $(5.6)$ and $(5.8)$,
one finds
$$\eqalignno{
&S'{}_D(\Phi',G')=S_D(\Phi,\mu,\bar\mu,A^*,A^{*\dagger};G)
={-1\over 12\pi}\int_\Sigma{d\bar z\wedge dz\over 2i}\tr\bigg\{
{1\over 1-\mu\bar\mu}\big(\bar\partial-\mu\partial_G-\ad A^*\big)\Phi&\cr&
{\exp\ad\Phi-1-\ad\Phi\over (\ad\Phi)^2}\big(\partial_G-\bar\mu\bar\partial
+\ad(GA^{*\dagger}G^{-1})\big)\Phi
-\bigg[F_G-\big(\partial_G-\bar\mu\bar\partial-(\bar\partial\bar\mu)\big)
{A^*\over 1-\mu\bar\mu}
&\cr&-\big(\bar\partial-\mu\partial_G-(\partial\mu)\big)
{GA^{*\dagger}G^{-1}\over 1-\mu\bar\mu}
+{[A^*,GA^{*\dagger}G^{-1}]\over 1-\mu\bar\mu}
\bigg]\Phi\bigg\}.&(7.13)\cr}$$
{}From here, the locality of $S'{}_D(\Phi',G')$ is apparent.

Let us compute $sS'{}_D(\Phi',G')$. The calculation can be performed in several
ways, the simplest of which is the following. Proceeding as done earlier
for ${\cal A}_{\rm ext}$, one finds that
$$s_{\rm diff}S'{}_D(\Phi',G')=0,\eqno(7.14)$$
so that, recalling that $sz'=s_{\rm diff}z'$,
$$\eqalignno{&sS'{}_D(\Phi',G')=(s-s_{\rm diff})S'{}_D(\Phi',G')=
{1\over 12\pi}\int_\Sigma{d\bar z'\wedge dz'\over 2i}
\tr\bigg[\Big((s-s_{\rm diff})\exp\Phi'\exp-\Phi'&\cr&
+\exp\ad\Phi'\big((s-s_{\rm diff})G'G'^{-1}\big)\Big)F'_{\exp\Phi'G'}
-(s-s_{\rm diff})G'G'^{-1}F'_{G'}\bigg].&(7.15)\cr}$$
Now, from $(7.7)$, using $(5.9)$, one has
$sH'H'^{-1}=s\exp\Phi'\exp-\Phi'+\exp\ad\Phi'(sG'G'^{-1})=
s_{\rm diff}H'H'^{-1}+W'_{H'}=s_{\rm diff}\exp\Phi'\exp-\Phi'+
\exp\ad\Phi'(s_{\rm diff}G'G'^{-1})+W'_{H'}$.
By comparing the two expressions, one finds that
$(s-s_{\rm diff})\exp\Phi'\exp-\Phi'
+\exp\ad\Phi'\big((s-s_{\rm diff})G'G'^{-1}\big)=W'_{H'}.$
By using this relation in $(7.15)$, recalling that $H'=\exp\Phi'G$
and using $(7.2)$, one finds
$$sS'{}_D(\Phi',G')=-{\cal A}'{}_{\rm ext}(W'_{H'},H')
-{1\over 12\pi}\int_\Sigma{d\bar z'\wedge dz'\over 2i}
\tr\bigg[\Big((s-s_{\rm diff})G'G'^{-1}F'_{G'}\bigg].\eqno(7.16)$$
{}From $(6.5)$, one finds easily that $sV=s_{\rm diff}V-MV$.
Recalling that $G'=V^{-1}GV^{-1\dagger}$ and $sG=0$, it is straightforward to
verify that $(s-s_{\rm diff})G'G'^{-1}=V^{-1}\big[M-c\Gamma_G
+G(M-c\Gamma_G)^\dagger G^{-1}\big]V$. From this relation, using $(2.6)$,
$(2.7)$, $(5.6)$ and $(5.8)$, $(7.16)$ can be written as
$$\eqalignno{&sS_D(\Phi,\mu,\bar\mu,A^*,A^{*\dagger};G)=
-{1\over 12\pi}\int_\Sigma{d\bar z\wedge dz\over 2i}
\tr\bigg\{\bigg[M-c\Gamma_G
+G(M-c\Gamma_G)^\dagger G^{-1}\bigg]\bigg[F_G&\cr&
-\big(\partial_G-\bar\mu\bar\partial-(\bar\partial\bar\mu)\big)
{A^*\over 1-\mu\bar\mu}
-\big(\bar\partial-\mu\partial_G-(\partial\mu)\big)
{GA^{*\dagger}G^{-1}\over 1-\mu\bar\mu}
+{[A^*,GA^{*\dagger}G^{-1}]\over 1-\mu\bar\mu}\bigg]\bigg\}&\cr&
-{\cal A}_{\rm ext}(W,W^\dagger,\Phi,\mu,\bar\mu,A^*,A^{*\dagger};G).
\vphantom{\int_\Sigma}&(7.17)\cr}$$
Both $(7.16)$ and $(7.17)$ show that the Weyl anomaly can be absorbed by
adding to the effective action $I'{}_{\rm aut}(H')$ a counterterm
$K S_D(\Phi,\mu,\bar\mu,A^*,A^{*\dagger};G)$ at the price
of creating a chirally non-split automorphism anomaly.

To obtain a chirally split automorphism anomaly, one has to find an other
counterterm analogous to $S_{VKLT}$. This is the Knecht-Lazzarini-Stora
action \ref{26}. Its expression is
$$\eqalignno{
&S_{KLS}(\mu,\bar\mu,A^*,A^{*\dagger};G,A,A^\dagger)
={-1\over 12\pi}\int_\Sigma{d\bar z\wedge dz\over 2i}
\tr\bigg\{(A-\Gamma_G)\bigg(A^*+{1\over 2}\mu(A-\Gamma_G)\bigg)&\cr&
+G\bigg[(A-\Gamma_G)\bigg(A^*+{1\over 2}\mu(A-\Gamma_G)
\bigg)\bigg]^\dagger G^{-1}+{1\over 1-\mu\bar\mu}\bigg[A^*GA^{*\dagger}G^{-1}
-{1\over 2}\bar\mu A^{*2}&\cr&-{1\over 2}\mu GA^{*\dagger2}G^{-1}\bigg]
\bigg\},&(7.18)\cr}$$
where $A$ is an arbitrary $(1,0)$ connection of $L$ such that
$$sA=0.\eqno(7.19)$$
By explicit calculation, using $(3.14)$ and $(6.21)$, one finds
$$\eqalignno{
&sS_{KLS}(\mu,\bar\mu,A^*,A^{*\dagger};G,A,A^\dagger)=
{1\over 12\pi}\int_\Sigma{d\bar z\wedge dz\over 2i}
\tr\bigg\{\bigg[M-c\Gamma_G
+G(M-c\Gamma_G)^\dagger G^{-1}\bigg]&\cr&\bigg[F_G
-\big(\partial_G-\bar\mu\bar\partial-(\bar\partial\bar\mu)\big)
{A^*\over 1-\mu\bar\mu}
-\big(\bar\partial-\mu\partial_G-(\partial\mu)\big)
{GA^{*\dagger}G^{-1}\over 1-\mu\bar\mu}
+{[A^*,GA^{*\dagger}G^{-1}]\over 1-\mu\bar\mu}\bigg]\bigg\}&\cr&
+\vphantom{\int}
{\cal A}_{\rm aut}(X_A,A^*{}_A;A)+\overline{{\cal A}_{\rm aut}(X_A,A^*{}_A;A)},
&(7.20)\cr}$$
where
$${\cal A}_{\rm aut}(X_A,A^*{}_A;A)=
{1\over 12\pi}\int_\Sigma{d\bar z\wedge dz\over 2i}
\tr\Big[X_A(\partial A^*{}_A-\bar\partial A+[A^*{}_A,A])\Big],\eqno(7.21)$$
with
$$A^*{}_A=A^*+\mu(A-\Gamma_G),\eqno(7.22)$$
$$X_A=X-C(A-\Gamma_G).\eqno(7.23)$$
$A^*{}_A$ and $X_A$ differ from $A^*$ and $X$ by a redefinition of the
reference $(1,0)$ connection entering in the definition of the latter from
$\Gamma_G$ to $A$.

The required counterterm contains two contributions. The first contribution
is given by $\kappa^\circ$ times the counterterm $\Delta I_{\rm cft}$
(cf. eq. $(4.7)$) and shifts the ordinary Weyl anomaly into the
chirally split diffeomorphism anomaly, as explained in sect. 4.
The second contribution is
given by $K$ times the counterterm
$$\Delta I(\Phi,\mu,\bar\mu,A^*,A^{*\dagger};G,A,A^\dagger)=
S_D(\Phi,\mu,\bar\mu,A^*,A^{*\dagger};G)
+S_{KLS}(\mu,\bar\mu,A^*,A^{*\dagger};G,A,A^\dagger),\eqno(7.24)$$
as follows from $(7.17)$ and $(7.20)$. Thus the extended Weyl invariant
effective action is
$$\eqalignno{
&I_{\rm ext}(\mu,\bar\mu,A^*,A^{*\dagger};{\cal R},\bar{\cal R},A,A^\dagger)=
I'{}_{\rm aut}(h^{\circ\prime},H')
+\kappa^\circ\Delta I_{\rm cft}(\phi^\circ,\mu,\bar\mu;
g^\circ,{\cal R},\bar{\cal R})&\cr&
+K\Delta I(\Phi,\mu,\bar\mu,A^*,A^{*\dagger};G,A,
A^\dagger),&(7.25)\cr}$$
which is easily verified to satisfy the Ward identity
$$\eqalignno{
&sI_{\rm ext}(\mu,\bar\mu,A^*,A^{*\dagger};{\cal R},\bar{\cal R},A,
A^\dagger)=\kappa^\circ{\cal A}_{\rm diff}(C,\mu;{\cal R})
+\kappa^\circ\overline{{\cal A}_{\rm diff}(C,\mu;{\cal R})}
&\cr&+K{\cal A}_{\rm aut}(X_A,A^*{}_A;A)
+K\overline{{\cal A}_{\rm aut}(X_A,A^*{}_A;A)}.&(7.26)\cr}$$
$I_{\rm ext}$ is manifestly independent from $g^\circ$ and $G$ by Weyl
invariance. Further, $I_{\rm ext}$ is holomorphically factorized \ref{26},
that is one has the chiral splitting $I_{\rm ext}(\mu,\bar\mu,A^*,
A^{*\dagger};{\cal R},\bar{\cal R},A,A^\dagger)=I_P(\mu,A^*;{\cal R},A)
+\overline{I_P(\mu,A^*;{\cal R},A)}$, where $I_P$ is a generalized
Polyakov action.

Before proceeding further,
a few remarks about the above results are in order. The
Donaldson action was originally introduced by Donaldson \ref{29} to study
the problem of the existence of Hermitian structures $(h^{\circ\prime},H')$
in a holomorphic vector bundle $E$ such that $F_H\sim h^\circ 1$. His method
generalizes the classical analysis of Liouville reducing the problem of
finding a constant curvature metric $h$ on a Riemann surface $\Sigma$
to a variational problem. The Donaldson action is formally similar
to a non compact gauged WZWN model. Indeed,
setting $\mu=0$ and $\bar\mu=0$ for simplicity, $(7.13)$ can be written as
$$\eqalignno{&S_D(\Phi,0,\bar 0,A^*,A^{*\dagger};G)=
S_D(\Phi,0,0,0,0;G)-{1\over 12\pi}\int_\Sigma{d\bar z\wedge dz\over 2i}
\bigg[-A^*\partial_G\exp\Phi\exp-\Phi
&\cr&+GA^{*\dagger}G^{-1}\bar\partial\exp-\Phi
\exp\Phi+A^*\exp\ad\Phi(GA^{*\dagger}G^{-1})-A^*GA^{*\dagger}G^{-1}\bigg].&
(7.27)\cr}$$
The `WZ field' is the Hermitian metric $H=\exp\Phi G$ and takes
values in the subset of Hermitian non singular elements of
$S(\Sigma,L\otimes\bar L)$. The `gauge fields' are $A^*$ and
$GA^{*\dagger}G^{-1}$. As the field $\Phi$ is determined uniquely
by $H$, $G$ and the Hermiticity condition $(5.10)$, the action is
manifestly singlevalued and no level quantization occurs.
Beyond the similarities, there are crucial differences. There
is no underlying group theoretic structure. The gauge fields are
not necessarily traceless and are related by Hermitian conjugation.
They are not yielded by gauging but by deformation of the holomorphic
structure of the bundle $E$.

It is interesting to find the anomalous Ward identity obeyed by
the generalized Polyakov action $I_P$ in differential form. This
consists of two parts:
$$\big(\bar\partial-\mu\partial_A-(\partial\mu)-\ad A^*\big)
{\delta I_P\over\delta A^*}(\mu,A^*;{\cal R},A)
={K\over 12\pi}\partial_AA^*,\eqno(7.28a)$$
$$\bigg[-\big(\bar\partial-\mu\partial-2(\partial\mu)\big)
{\delta I_P\over\delta\mu}+\tr\bigg(\partial_AA^*{\delta I_P
\over\delta A^*}\bigg)\bigg](\mu,A^*;{\cal R},A)
={\kappa^\circ\over 12\pi}\big(\partial^3
+2{\cal R}\partial+(\partial{\cal R})\big)\mu.\eqno(7.28b)$$
$(7.28b)$ is the counterpart of the conformal Ward identity, but it
differs from the latter by the second term
within square brackets \ref{39}.

As shown extensively in sects. 5 and 6,
the holomorphic and Hermitian geometry of $E$ reduce essentially to
those of $\det E$ and $\proj E$. This is reflected in the splitting of the
fields $A^*$ and $\Phi$ in their trace parts $a^*$ and $\phi$ and traceless
parts $\hat A^*$ and $\hat\Phi$. Correspondingly, the action of the
symmetry group factorizes into an action on $\det E$ and one on $\proj E$.
This again is reflected in the splitting of the ghost fields $M$ and $W$ in
their trace parts $m$ and $w$ and traceless parts $\hat M$ and $\hat W$.
One may express all the anomalies and the counterterm given
above in terms of such component fields. As turns out, each of such
functionals results to be a sum of two terms, the first depending only on
the trace part fields and  the second depending only on the traceless
part fields, and thus referring to $\det E$ and $\proj E$, respectively.
The trace part term of each functional appears to have the same form
as that of the corresponding functional in ordinary conformal field theory.
Thus, it will contribute to the ordinary Weyl or diffeomorphism anomaly.
Let us illustrate this in greater detail.

Assume first that the normalized weight $j$ of $E$ is non-zero. Then, for
any Hermitian structure $(h^{\circ\prime},H')$ of $E$ subordinated to
$\sans A$ the field locally defined by
$$h'{}_{\rm ind}=(\det H')^{1/jr}\eqno(7.29)$$
is an induced Hermitian metric of $\Sigma$ with respect to $\sans a$
provided one restricts oneself to the reduction ${\sans A}'_0$
of $\sans A$ (cf. sect. 5). I shall stick to such reduction for the
rest of this section. The Donaldson parametrization of the Hermitian
structures of $E$ yields directly a Liouville parametrization of
the metrics $h'{}_{\rm ind}$ with respect
to the reference metric $g_{\rm ind}=(\det G)^{1\over jr}$. From $(5.9)$
and $(5.14)$, it follows that the Liouville field $\phi_{\rm eff}$
of $h'{}_{\rm ind}$ is
$$\phi_{\rm eff}=\phi/j+\varphi_\sigma,\eqno(7.30a)$$
where
$$\varphi_\sigma=2{\rm Re}\hskip1pt\sigma/j.\eqno(7.30b)$$
$\varphi_\sigma$ is a real valued element of $S(\Sigma,1)$,
since $\exp(-pr\sigma)$ is a section of the $1$--cocycle
$e\otimes e'^{-1}$ and the latter is flat unitary when restricted
to the reductions ${\sans A}_0$ and ${\sans A}'_0$.
Note that the indetermination of $\sigma$ modulo $2\pi i/pr$ times
integers does not affect $\varphi_\sigma$. From $(7.30)$ and $(6.12)$,
one can obtain easily an expression for $s\varphi_\sigma$. Introducing the
ghost field
$$w_{1{\rm ind}}={\rm Re}\hskip1pt w/j,\eqno(7.31)$$
one finds that $s\phi_{\rm eff}$ is given by $(3.18)$ with $\phi$
and $w_1$ replaced by $\phi_{\rm eff}$ and
by $w_{1{\rm ind}}$. Thus, $\phi_{\rm eff}$ rather than $\phi$
appears to be the proper counterpart of the Liouville field in
the present context.

By a straightforward calculation, one finds
$$\eqalignno{
&{\cal A}_{\rm ext}(W,W^\dagger,\Phi,\mu,\bar\mu,A^*,A^{*\dagger};G)
=j^2r{\cal A}_{\rm conf}(w_{1{\rm ind}},\phi_{\rm eff},\mu,\bar\mu;g_{\rm ind})
&\cr&+{\cal A}_{\rm ext}(\hat W, \hat W^\dagger,\hat\Phi,\mu,\bar\mu,\hat A^*,
\hat A^{*\dagger};G),&(7.32)\cr}$$
where the first term is given explicitly by $(4.2)$.
Thus, the Ward identity $(7.1)$ may be written as
$$\eqalignno{&sI'{}_{\rm aut}(h^{\circ\prime},H')=
\kappa^\circ{\cal A}_{\rm conf}(w_1,\phi^\circ,\mu,\bar\mu;g^\circ)
+Kj^2r{\cal A}_{\rm conf}(w_{1{\rm ind}},\phi_{\rm eff},\mu,\bar\mu;
g_{\rm ind})&\cr&+K{\cal A}_{\rm ext}(\hat W, \hat W^\dagger,\hat\Phi,\mu,
\bar\mu,\hat A^*,\hat A^{*\dagger};G).&(7.33)\cr}$$
Note further that the effective Liouville
field $\phi_{\rm eff}$ is a non local functional of $\mu,\bar\mu,A^*,
A^{*\dagger}$, for $\sigma$ is a non local functional of $\mu,A^*$. However,
the induced Weyl anomaly itself is local.

The Donaldson action and the Knecht-Lazzarini-Stora action undergo a
similar decomposition.
$$\eqalignno{&S_D(\Phi,\mu,\bar\mu,A^*,A^{*\dagger};G)
=j^2r\big[S_L(\phi_{\rm eff},\mu,\bar\mu;g_{\rm ind})
-S_L(\varphi_\sigma,\mu,\bar\mu;g_{\rm ind})\big]&\cr&
+S_D(\hat\Phi,\mu,\bar\mu,\hat A^*,\hat A^{*\dagger};G),&(7.34)\cr}$$
where the Liouville action is given by $(4.4)$.
$$\eqalignno{&S_{KLS}(\mu,\bar\mu,A^*,A^{*\dagger};G,A,A^\dagger)
=j^2r\big[S_{VKLT}(\mu,\bar\mu;g_{\rm ind},{\cal R},
\bar{\cal R})+S_L(\varphi_\sigma,\mu,\bar\mu;g_{\rm ind})&\cr&
-\widehat{\Delta I}(\mu,\bar\mu,A^*,A^{*\dagger};
g_{\rm ind},a,\bar a,{\cal R},\bar{\cal R})\big]
+S_{KLS}(\mu,\bar\mu,\hat A^*,\hat A^{*\dagger};G,\hat A
+j\gamma_{g_{\rm ind}}1,\hat A^\dagger+j\bar\gamma_{g_{\rm ind}}1),
&\cr&&(7.35)\cr}$$
where $\cal R$ is an $\sans a$--holomorphic projective connection,
$S_{VKLT}$ is given by $(4.5)$ and
$$\eqalignno{
&\widehat{\Delta I}(\mu,\bar\mu,A^*,A^{*\dagger};g,a,\bar a,{\cal R},
\bar{\cal R})=
{-1\over 12\pi}\int_\Sigma{d\bar z\wedge dz\over 2i}\bigg\{
\mu\bigg[\partial\big(\rho_a+\chi_\sigma\big)
-{1\over 2}\big(\rho_a+\chi_\sigma\big)^2-{\cal R}\bigg]&\cr&
+(\rho_a-\gamma_g)\chi^*_\sigma
+\bar\mu\bigg[\bar\partial\big(\bar\rho_a+\bar\chi_\sigma\big)
-{1\over 2}\big(\bar\rho_a+\bar\chi_\sigma\big)^2-\bar{\cal R}\bigg]
+(\bar\rho_a-\bar\gamma_g)\bar\chi^*_\sigma
+(1/2)\chi_\sigma\chi^*_\sigma
&\cr&+(1/2)\bar\chi^*_\sigma\bar\chi_\sigma
-\partial\bar\partial\ln g\varphi_\sigma\bigg\},&(7.36a)\cr}$$
with
$$\chi_\sigma=\partial\sigma/j,\quad\chi^*_\sigma=\bar\partial\sigma/j,
\eqno(7.36b)$$
$$\rho_a=a/j.\eqno(7.36c)$$
Because of the special choice of trivializations made, $\chi_\sigma$
and $\chi^*_\sigma$ are conformal fields of weights $1,0$ and $0,1$,
respectively. Further, the indetermination of $\sigma$ modulo
$2\pi i/pr$ times integers does not affect them.
Hence, the above functional is well-defined, the integrand being a
globally defined conformal field of weights $1,1$. Note that
$\widehat{\Delta I}$ is a non local functional of
$\mu,\bar\mu,A^*,A^{*\dagger}$, for $\sigma$ is a non local
functional of $\mu,A^*$. It is however chirally split, since $\sigma$
is holomorphic.

By combining $(7.24)$, $(7.34)$ and $(7.35)$, one finds
$$\eqalignno{
&\Delta I(\Phi,\mu,\bar\mu,A^*,A^{*\dagger};G,A,A^\dagger)+
j^2r\widehat{\Delta I}(\mu,\bar\mu,A^*,A^{*\dagger};g_{\rm ind},a,\bar a,
{\cal R},\bar{\cal R})&\cr&
=j^2r\Delta I_{\rm cft}(\phi_{\rm eff},\mu,\bar\mu;g_{\rm ind},
{\cal R},\bar{\cal R})
+\Delta I(\hat\Phi,\mu,\bar\mu,\hat A^*,\hat A^{*\dagger};G,\hat A
+j\gamma_{g_{\rm ind}}1,\hat A^\dagger+j\bar\gamma_{g_{\rm ind}}1).
&\cr&&(7.37)\cr}$$
Note that the right hand side is the sum of the counterterm
$\Delta I_{\rm cft}$ encountered in ordinary conformal field theory
plus the projective part of the counterterm $\Delta I$. The strange looking
connection $\hat A+j\gamma_{g_{\rm ind}}1$ is in fact what is needed
since $A$ appears always in the combination $A-\Gamma_G$ (cf. eq. $(7.18)$)
while the projective part of $\Delta I$ should be independent from
$\det G$.

A straightforward calculation using $(6.26)$ shows that
$$\eqalignno{
&j^2rs\widehat{\Delta I}(\mu,\bar\mu,A^*,A^{*\dagger};g_{\rm ind},
a,\bar a,{\cal R},\bar{\cal R})
=j^2r\big[{\cal A}_{\rm diff}(C,\mu;{\cal R})
+\overline{{\cal A}_{\rm diff}(C,\mu;{\cal R})}\big]
&\cr&-\big[{\cal A}_{\rm aut}(x1,a^*1;a1)
+\overline{{\cal A}_{\rm aut}(x1,a^*1;a1)}\big].
&(7.38)\cr}$$
Thus, defining
$$\eqalignno{
&\widehat{I_{\rm ext}}(\mu,\bar\mu,A^*,A^{*\dagger};{\cal R},\bar{\cal R},
A,A^\dagger)
=I_{\rm ext}(\mu,\bar\mu,A^*,A^{*\dagger};{\cal R},\bar{\cal R},A,
A^\dagger)&\cr&
+Kj^2r\widehat{\Delta I}(\mu,\bar\mu,A^*,A^{*\dagger};g_{\rm ind},a,\bar a,
{\cal R},\bar{\cal R}),&(7.39)\cr}$$
one has
$$\eqalignno{&s\widehat{I_{\rm ext}}(\mu,\bar\mu,A^*,A^{*\dagger};{\cal R},
\bar{\cal R},A,A^\dagger)
=\big(\kappa^\circ+Kj^2r\big)\big[{\cal A}_{\rm diff}(C,\mu;{\cal R})
+\overline{{\cal A}_{\rm diff}(C,\mu;{\cal R})}\big]
&\cr&+K\big[{\cal A}_{\rm aut}(\hat X,\hat A^*;\hat A)
+\overline{{\cal A}_{\rm aut}(\hat X,\hat A^*;\hat A)}\big].&(7.40)\cr}$$
As $I_{\rm ext}$, $\widehat{I_{\rm ext}}$ does not depend on $g^\circ$ and
$G$ by extended Weyl invariance. In fact, it is easy to check from
$(7.36a)$ that $\widehat{\Delta I}$ does not depend on the scale of the
metric $g$.

The above results need to be adequately discussed and interpreted.
{}From the Ward identity $(7.33)$, it appears
that the extended Weyl anomaly induces an ordinary Weyl anomaly with
induced central charge $Kj^2r$. The anomaly expresses
the non trivial dependence on the scale of the metric $h'{}_{\rm ind}$
in the effective action and is thus distinguished from the ordinary Weyl
anomaly referring to the metric $h^\circ$ with central charge $\kappa^\circ$.
However, if one restricts oneself to special deformations and Hermitian
structures of $E$ (cf. sect. 5) and reduces the symmetry to
the special symmetry (cf. sect. 6), then $g_{\rm ind}=g^\circ$,
$\phi_{\rm eff}=(1/j)\phi=\phi^\circ$ and $w_{1{\rm ind}}=w_1$,
as follows from combining $(5.16)$, $(5.18)$, $(5.19)$ and $(6.28)$ with
$(7.30)$ and $(7.31)$. Thus, the two Weyl anomalies coalesce into a single
one with central charge $\kappa^\circ+Kj^2r$.

As shown above,
upon modifying the effective action $I_{\rm ext}$ by adding the functional
$K\widehat{\Delta I}$, one obtains a new effective action
$\widehat{I_{\rm ext}}$
obeying the anomalous Ward identity $(7.40)$. The automorphism anomaly is
still chirally split. However, $\widehat{\Delta I}$ is a non local functional.
So, the anomalies $(7.26)$ and $(7.40)$ are not equivalent as classes of the
local $s$ cohomology. However, if one restricts oneself once more
to special deformations and Hermitian structures
and reduces the symmetry to the special symmetry,
then $\phi_{\rm eff}=\phi^\circ$, as shown in the previous paragraph,
and further $\varphi_\sigma=0$, $\chi_\sigma=0$ and $\chi^*_\sigma=0$ by
$(5.16)$, $(7.30)$ and $(7.36b)$. The functional $\widehat{\Delta I}$
becomes in this way local, as appears from $(7.36a)$ by inspection.
The anomalies $(7.26)$ and $(7.40)$ are then equivalent in the reduced local
$s$ cohomology associated to the special symmetry. The central charge
is again $\kappa^\circ+Kj^2r$.

To summarize, I have shown that restricting to special holomorphic and
Hermitian structures and their deformations and reducing the symmetry to the
special symmetry shifts the ordinary central charge. The effective
central charge is given by
$$\kappa_{\rm eff}=\kappa^\circ+\kappa_{\rm ind}=\kappa^\circ+Kj^2r.
\eqno(7.41)$$

In the above calculations, I assumed that $j\not=0$ since the definition
of various fields involved a factor $1/j$. The analysis would seem to break
down when $j=0$. Happily, this is not so, provided one restricts oneself
from the beginning to special deformations and Hermitian structures
and to the special symmetry group. This follows from a straightforward
computation based on the relations $(5.15)$, $(5.16)$, $(5.18)$, $(5.19)$,
$(6.27)$ and $(6.28)$. The upshot is that relations $(7.32)$, $(7.33)$,
$(7.34)$, $(7.35)$, $(7.37)$ and $(7.40)$ hold with the terms proportional
to $j^2r$ formally set to zero. In the case considered, there is no induced
central charge so that $\kappa_{\rm eff}=\kappa^\circ$.

The above discussion provides shows convincingly the naturality of restricting
to special structures and special symmetry. Henceforth, this will be assumed
tacitly unless otherwise stated.

Because only special deformations are admitted, $\widehat{I_{\rm ext}}$
is effectively a functional of $\mu$, $\bar\mu$, $\hat A^*$ and
$\hat A^{*\dagger}$ (cf. sect. 5). Since $\widehat{\Delta I}$ is chirally
split, $\widehat{I_{\rm ext}}$ satisfies the holomorphic factorization
theorem if $I_{\rm ext}$ does. Thus, one has a relation of the form
$\widehat{I_{\rm ext}}(\mu,\bar\mu,\hat A^*,\hat A^{*\dagger}
;{\cal R},\bar{\cal R},A$, $A^\dagger)=\widehat{I_P}(\mu,\hat A^*;
{\cal R},A)$ $+$ $\overline{\widehat{I_P}(\mu,\hat A^*;
{\cal R},A)}$, where $\widehat{I_P}$ is a generalized Polyakov action.
$\widehat{I_P}$ satisfies a Ward identity of the same type as $(7.28)$
with $A^*$ replaced by $\hat A^*$ and $\kappa^\circ$
substituted by $\kappa_{\rm eff}$.

Being a functional of $\mu$ and $\hat A^*$, $\widehat{I_P}$ depends on
the holomorphic geometry of $\proj E$ rather than that of $E$ (cf sect. 5).
This is a rather non trivial geometric property of the whole construction.

At this point, it is worthy considering an example. The following model
is in fact a variant of the one originally envisaged in ref. \ref{26}
and may considered as a generalization of the bosonic $b$--$c$ system.
For any vector bundle $E$ and any holomorphic structure $\sans A$,
the classical action is
$$S(\Psi^\vee,\Psi)={1\over\pi}\int_{\Sigma}{d\bar z\wedge dz\over 2i}
\Psi^\vee\bar\partial\Psi,\eqno(7.42)$$
where $\Psi$ and $\Psi^\vee$ vary, respectively, in
$S(\Sigma, k^{\otimes{1\over 2}}\otimes L)$ and
$S(\Sigma, k^{\otimes{1\over 2}}\otimes L^\vee)$. For such model,
it is possible to compute the central charges $\kappa^\circ$ and
$K$:
$$\kappa^\circ=-r,\quad K=12,\eqno(7.43)$$
\ref{26}. From here, one can compute $\kappa_{\rm eff}$ by using $(7.41)$.
The result can be written as
$$\kappa_{\rm eff}=(12j^2-1)r=2r(6\tilde\jmath^2-6\tilde\jmath+1), \quad
\tilde\jmath=1/2+j.\eqno(7.44)$$
$\tilde\jmath$ is the effective conformal weight of the field $\Psi$.
The two contributions to $\tilde\jmath$ correspond respectively to
the first and second tensor factor of the $1$--cocycle
$k^{\otimes{1\over 2}}\otimes L$. Note that the second form of the
expression of $\kappa_{\rm eff}$ is that of $r$ copies of a spin $\tilde\jmath$
bosonic $b$--$c$ system.
\vskip.6cm
\item {8.} {\bf Flat vector bundles and their flat structures
and extended 2D gravity.}
\vskip.4cm
\par
Within the category of complex vector bundle, the flat ones have
a considerable salience in applications. By definition, flat vector
bundles admit flat structures. Recall that a flat structure $\sans F$
subordinated to a holomorphic structure $\sans A$ of a flat vector bundle
$E$ is a maximal reduction of $\sans A$ such that $L_{ab}$ is a constant
matrix for all pairs $a,b$ of overlapping trivializations contained in
$\sans F$. I shall write $({\sans A},{\sans F})$ to emphasize
that $\sans F$ is subordinated to $\sans A$. One may consider several
flat structures $({\sans A}',{\sans F}')$, $({\sans A}'',{\sans F}'')$
viewed as deformations of a reference structure $({\sans A},{\sans F})$.
The problem arises of finding a suitable parametrization of the family
of such deformations. Proceeding as in sect. 5, one finds that the
deformations $({\sans A}',{\sans F}')$ are parametrized in one-to-one
fashion by the triples $(\mu,A^*,B)$, where $(\mu,A^*)$ specify the
deformation ${\sans A}'$ and $B$ is given by
$$B_a=(\partial_a-A_a)V_{aa'}V_{aa'}{}^{-1}, \quad a'\in{\sans F}'.
\eqno(8.1)$$
$B$ is an element of $S(\Sigma,k\otimes{\rm End}\hskip1pt L)$.
$\mu$, $A^*$ and $B$ satisfy the identity
$$\bar\partial(A+B)-\partial(A^*+\mu B)+[A+B,A^*+\mu B]=0,
\eqno(8.2)$$
which is the zero curvature condition for the connection
$(A+B)dz+(A^*+\mu B)d\bar z$. This relation can be cast also as
$$\big(\bar\partial-\mu\partial_A-(\partial\mu)-\ad A^*\big)B
=\partial_AA^*.\eqno(8.3)$$
{}From here, it can be seen that the flat structures ${\sans F}'$ subordinated
to the same holomorphic structure ${\sans A}'$ are in one-to-one
correspondence the ${\sans a}'$--holomorphic elements of $S(\Sigma,
k'\otimes{\rm End}\hskip1pt L')$ (the $0$--th sheaf cohomology of the
${\sans a}'$--holomorphic $1$--cocycle $k'\otimes{\rm End}\hskip1pt L'$).
It follows from $(6.4)$ that the automorphism
group of $E$ maps flat structures into flat structures. From $(3.14)$,
$(3.15)$, $(6.20)$, $(6.21)$ and $(6.22)$ one finds the relations
$$s(A+B)=\partial(CB-X)-[A+B,CB-X],\eqno(8.4)$$
$$s(A^*+\mu B)=\bar\partial(CB-X)-[A^*+\mu B,CB-X],\eqno(8.5)$$
$$s(CB-X)=(CB-X)^2.\eqno(8.6)$$
One may consider also special deformations of a flat structure
for which $\det V_{aa'}=1$. Such deformations are parametrized in
one-to-one fashion by the triples $(\mu,A^*,B)$ such that $A^*$ and
$B$ are traceless (assuming that $\tr A=0$). In that
instance, I shall revert to the familiar notation $\hat A^*$ and
$\hat B$.
For special automorphisms, one has similarly that $X$ is traceless,
so that $X=\hat X$ (cf. sect. 6).

In concrete applications to extended $2D$ gravity, the vector bundle
$E$ is of the form $\epsilon^j\otimes\hat E$ with $j\in{\bf Z}/2$, where
$\hat E$ is a flat vector bundle. To any holomorphic structure
$hat{\sans A}$ of $\hat E$, there is
associated a holomorphic structure $\sans A$ of $E$ such that
${\sans a}=\hat{\sans a}$ and $L_{ab}=k^{\otimes m}{}_{ab}\hat L_{ab}$.
The trivializations $(z_a,u_a)$ of $\sans A$ are of the form
$v_a\otimes\hat u_a$ for sum trivialization $(\hat z_a,\hat u_a)$,
where $z_a=\hat z_a$ and $v_a$ is the trivialization of
$\epsilon^{\otimes m}$ corresponding to $\hat z_a$. Since
$S(\Sigma,\bar k\otimes{\rm End}\hskip1pt L)=
S(\Sigma,\bar k\otimes{\rm End}\hskip1pt\hat L)$, there is a one-to-one
correspondence between the special deformations of $\hat E$ and $E$,
the corresponding deformations being described by the same pair of
$(\mu,\hat A^*)$ of geometrical fields.
Similarly, there is an isomorphism of the special automorphism
group of $\hat E$ onto that of $E$ for a given $\mu$, defined by
$\alpha_{ba}=\varpi_{ab}(f_\alpha;\mu)^{-2m}\hat\alpha_{ba}$.

Restricting to special deformations and
special symmetry, the generalized Polyakov action $\widehat{I_P}$ is a
functional of $\mu$ and $\hat A^*$ obeying a Ward identity of the form
$(7.28)$ with $A^*$ replaced by $\hat A^*$. If one sets $B=(12\pi/K)
\delta\widehat{I_P}/\delta A^*$, $(7.28a)$ has the same form
as $(8.3)$. The Ward identity assumes then the form of a zero
curvature condition, a recurrent
fact in all extensions of $2D$ gravity. In more geometrical terms,
$\widehat{I_P}$ defines a section of the family of flat structures of
$\hat E$ viewed as a space fibered over the set of holomorphic
structures. By substituting $(7.28a)$ in $(7.28b)$,
one finds that $P=(12\pi/\kappa_{\rm eff})
[-\delta\widehat{I_P}/\delta\mu+(6\pi/K)\tr(\delta\widehat{I_P}/
\delta A^*)^2]+{\cal R}$ satisfies
$(\bar\partial-\mu\partial-2(\partial\mu))P=\partial^3\mu$,
which characterizes $P$ as an ${\sans a}'$--holomorphic projective
connection \ref{37,39}. So $\widehat{I_P}$ yields a section of the
space of projective coordinate structures of $\Sigma$ viewed as a fibered
space over the set of holomorphic structures as in ordinary conformal
field theory \ref{39}.
\vskip.6cm
\item {9.} {\bf Applications to extended 2D gravity.}
\vskip.4cm
\par
In this final section, I shall describe briefly a few examples of the
general framework expounded in the previous sections. I shall consider
vector bundles $E$ of the form $\epsilon^j\otimes\hat E$ with $j\in{\bf Z}/2$,
where $\hat E$ is a flat vector bundle. There is a choice of $\hat E$ which
relevant for applications which now I shall introduce. The construction
exploits $A_1$ embeddings into simple Lie algebras and is in line with
recent work in extended $2D$ gravity in which $A_1$ embeddings
play a prominent role \ref{33-34}.
\vskip.4cm
$a)$ {\it The Drinfeld-Sokolov vector bundle.}
\vskip.4cm
Let $\sans g$ be a simple complex Lie algebra. Consider a $A_1$ embedding into
$\sans g$. This is given by a triple of Lie algebra elements $t_{-1},t_0,
t_{+1}$ with Lie brackets
$$[t_{+1},t_{-1}]=2t_0,\quad [t_0, t_{\pm 1}]=\pm t_{\pm 1},\eqno(9.1)$$
the other brackets being zero.
Chosen any representation $R$ of $\sans g$, one may define
$$\hat L_{ab}=k_{ab}{}^{-t_0}\exp(\partial_ak_{ab}{}^{-1}t_{-1})\eqno(9.2)$$
on a Riemann surface $\Sigma$ with holomorphic structure $\sans a$
\footnote{${}^9$}{Since I will work with an arbitrary but fixed
representation $R$, I shall not distinguish notationally between Lie
algebra elements and their representatives.}.
The collection $\{k_{ab}{}^{-t_0}\}$ defines a ${\rm SL}(r,{\rm C})$--valued
$1$--cocycle, where $r=\dim R$, since $R$ induces a representation
of the $A_1$ embedding and this is equivalent to a unitary one in which
$t_0$ is represented by a diagonal matrix with half-integer entries.
By using this remark, it is straightforward to verify that $\hat L$
is a holomorphic ${\rm SL}(r,{\bf C})$ $1$--cocycle.  One may also
easily verify that $\hat L$ is unstable \ref{36}. Indeed, the fields of the
form $Z=\psi t_{-1}$, where $\psi$ is a holomorphic differential of
$S(\Sigma,k)$, are non trivial holomorphic elements of
$S(\Sigma,{\rm End}\hskip1pt\hat L)$. $\hat L$ possesses a distinguished
$(1,0)$ holomorphic connection
$$\hat A=(1/2)t_{+1}-{\cal R} t_{-1},\eqno(9.3)$$
where $\cal R$ is a holomorphic projective connection. This in turn
shows the flatness of $\hat L$ \ref{36}.

The $1$--cocycle $\hat L$ defines a smooth flat vector bundle
$DS(t,R)$, which will be called the Drinfeld--Sokolov bundle of
the $A_1$ embedding $t$ into $\sans g$ in the representation
$R$, because of the special form of the connection $\hat A$
\ref{42}. It further provides a holomorphic structure
$\sans A$ naturally associated to the underlying
holomorphic structure $\sans a$ of $\Sigma$, by $(9.2)$.
Hence, any deformation ${\sans a}'$ of $\sans a$ induces a
deformation ${\sans A}'$ of $\sans A$. The associated deformation
intertwiners are the intertwiner $\lambda$ of ${\sans a}'$ and
$$\hat V_{aa'}=\lambda_{aa'}{}^{-t_0}
\exp(\partial_a\lambda_{aa'}{}^{-1}t_{-1}).\eqno(9.4)$$
Hence, the deformation considered is special. The geometric field
describing the deformation are the Beltrami differential
$\mu$ and the  gauge field
$$\hat A^*=(\mu/2) t_{+1}-\partial\mu t_0-(\partial^2+{\cal R})\mu t_{-1},
\eqno(9.5)$$
where the chosen reference connection $\hat A$ entering into
the definition of $\hat A^*$ according to $(5.6)$ is given by
$(9.3)$.

The action of the diffeomorphism group on the deformations
${\sans a}'$ of $\sans a$ can be lifted to an action onto the induced
deformations ${\sans A}'$ of $\sans A$ constructed above. This leads
to a homomorphism of the diffeomorphism group of $\Sigma$ into the
special automorphism group of $DS(t,R)$.
The action of the Slavnov operator on $\mu$ and $\hat A^*$
is given by $(3.14)$ and $(6.21)$ with
$$\hat X=-(C/2) t_{+1}+\partial C t_0
+(\partial^2+{\cal R})C t_{-1},\eqno(9.6)$$
whose validity is restricted to the image of the lift in the symmetry
Lie algebra.

It is interesting to compute the projective part of the automorphism
anomaly ${\cal A}_{\rm aut}$ for the lift of the diffeomorphism
group. By a straightforward calculation, one finds
$${\cal A}_{\rm aut}(\hat X,\hat A^*;\hat A)
=\tr_R(t_0{}^2){\cal A}_{\rm diff}(C,\mu;{\cal R}),\eqno(9.7)$$
where $\tr_R$ is the trace in the representation $R$ of the Lie algebra
$\sans g$. This yields a further contribution $\kappa_{\rm lift}
=\tr_R(t_0{}^2)$ to the central charge of the form of an improvement
term. On account of $(7.41)$, the total central charge is
$$\kappa_{\rm tot}=\kappa_{\rm eff}+\kappa_{\rm lift}=
\kappa^\circ+K\big(j^2\dim R+\tr_R(t_0{}^2)\big).\eqno(9.8)$$

As an application, consider the system $(7.42)$ in the case where
$E=\epsilon^j\otimes DS(t,{\sans g})$.
By combining $(7.44)$ and $(9.8)$, one finds
$$\kappa_{\rm tot}=\tr_R\big[2(6\tilde\jmath^2-6\tilde\jmath+1)1+t_0{}^2
\big],\quad\tilde\jmath=1/2+j.\eqno(9.9)$$
\vskip.4cm
$b)$ {\it W gravity in the light cone gauge.}
\vskip.4cm
Next, one would like to see whether the above framework can accommodate
well-known models of extended $2D$ gravity, in particular $W$ gravity.
To this end, I shall consider a class of flat structures of the bundle
$DS(t,{\sans g})$ defined by the constraint
$$\ad t_{-1}\hat B=0,\eqno(9.10)$$
where $\hat B$ is given by $(8.1)$ with $\hat A$ given by $(9.3)$.
By combining $(8.2)$ and $(9.10)$, one obtains then
$$\ad t_{-1}\big[\partial(\hat A^*+\mu\hat B)
-[\hat A+\hat B,\hat A^*+\mu\hat B]\big]=0.\eqno(9.11)$$
When restricted to the residual gauge symmetry left over by the
constraint, the ghost field $\hat X$ obeys the following relation
$$\ad t_{-1}\big[\partial(C\hat B-\hat X)
-[\hat A+\hat B,C\hat B-\hat X]\big]=0,\eqno(9.12)$$
as follows from $(8.4)$ and $(9.10)$. Applying the Slavnov
operator $s$ on the expressions defined by the left hand sides
of eqs. $(9.11)$--$(9.12)$ and using $(8.2)$, $(8.4)$--$(8.6)$
leads to linear combinations of those very same expressions and their
derivatives. Thus there are no more constraints besides the ones
already indicated. It is not difficult to verify that
$(9.10)$--$(9.12)$ are compatible with changes of trivializations.
The solution $(9.11)$--$(9.12)$ can be found as follows.

The adjoint representation of the $A_1$ embedding $t_{-1}$, $t_0$,
$t_{+1}$ in $\sans g$ is reducible. Let us denote
by $\Pi$ the set of the representations of $A_1$ appearing in the
decomposition each counted with its multiplicity. To the
representation $\eta$ of spin $l_\eta$, there is associated a
distinguished set of generators $t_{\eta,m_\eta}$, $m_\eta
=-l_\eta,-l_\eta+1,\cdots,l_\eta-1,l_\eta$ of $\sans g$ such that
$$[t_d,t_{\eta,m_\eta}]=G^d{}_{l_\eta,m_\eta}t_{\eta,m_\eta+d},
\quad d=-1,0,+1,\eqno(9.13a)$$
where
$$G^{\pm 1}{}_{l,m}=[l(l+1)-m(m\pm 1)]^{1\over 2},\quad G^0{}_{l,m}=m.
\eqno(9.13b)$$
The Lie brackets of the $t_{\eta,m_\eta}$'s have the following form
$$[t_{\eta,m_\eta},t_{\zeta,m_\zeta}]=
\sum_{\xi\in\Pi}\sum_{m_\xi=-l_\xi}^{l_\xi}F_{\eta,\zeta}{}^\xi
(l_\eta,m_\eta;l_\zeta,m_\zeta | l_\xi,m_\xi)t_{\xi,m_\xi},
\eqno(9.14)$$
where $(l_\eta,m_\eta;l_\zeta,m_\zeta | l_\xi,m_\xi)$ is a
Clebsch-Gordan coefficient and the $F_{\eta,\zeta}{}^\xi$'s are imaginary
constants depending only on the $A_1$ embedding $t$. There always is an
element of $o\in\Pi$ such that $l_o=1$, since the $t_{-1},t_0,t_{+1}$
themselves span a representation of $A_1$ into $\sans g$ with
$t_{o,\pm 1}=\mp 2^{-{1\over 2}}t_{\pm 1}$ and $t_{o,0}=t_0$.
Below, I shall restrict to $A_1$ embeddings for which $\Pi$
contains only integers spins. The simpleness of $\sans g$ and the
non degeneracy of the Cartan form imply that $\Pi$ cannot contain
spin $0$ representations. One has further
$$\tr_R\big(t_{\eta,m_\eta}t_{\zeta,-m_\zeta}\big)=N_\eta
(-1)^{l_\eta-m_\eta}\delta_{\eta,\zeta}\delta_{m_\eta,m_\zeta},
\eqno(9.15)$$
where $N_\eta$ is a normalization constant.
In practice, the most used $A_1$ embedding is the so-called principal
$A_1$, for which $t_0$ is the dual Weyl vector $\rho^\vee$.
In that case, each element of $\Pi$ is non degenerate except for
$l=2k-1 $ of the algebra $D_{2k}$ which is doubly degenerate.

The solution of the constraint $(9.10)$ can be written in the form
$$\hat B=\sum_{\eta\in\Pi}\psi_\eta t_{\eta,-l_\eta},\eqno(9.16)$$
where the $\psi_\eta$ are elements of $S(\Sigma,k^{\otimes l_\eta+1})$.
The solution of the constraint $(9.11)$ can be found by means of techniques
analogous to those worked out in the second reference \ref{23}.
Let $Q$ be the operator defined by $Qt_{\eta,m_\eta}=
2(G^{+1}{}_{l_\eta,m_\eta})^{-2}t_{\eta,m_\eta}$ for $m_\eta<l_\eta$ and
$Qt_{\eta,l_\eta}=0$ and let $l_{\rm max}$ be the largest spin
in $\Pi$. Then, the result can be written as
$$\hat A^*=-\mu\hat B
+\Big[1+Q\ad t_{-1}(\partial_{\hat A}-\ad\hat B)\Big]^{2l_{\rm max}+1}
\sum_{\eta\in\Pi}\nu_\eta t_{\eta,l_\eta},\eqno(9.17)$$
where the $\nu_\eta$ are fields from $S(\Sigma, \bar k\otimes
k^{\otimes-l_\eta})$ and may be viewed as generalized Beltrami
fields. Similarly, solving $(9.12)$ yields
$$\hat X=C\hat B
-\Big[1+Q\ad t_{-1}(\partial_{\hat A}-\ad\hat B)\Big]^{2l_{\rm max}+1}
\sum_{\eta\in\Pi}y_\eta t_{\eta,l_\eta},\eqno(9.18)$$
where $y_\eta=-\hat X_{\eta,l_\eta}$ is a section of $k^{\otimes-s_\eta}
\otimes{\rm Lie}\hskip1pt\big({\rm Aut}_c(DS(t,R))
\times{\rm Weyl}(DS(t,R))\big)^\vee$.

The $\psi_\eta$'s and $\nu_\eta$'s are not independent, since they
must satisfy a set of differential equations following from
$(8.3)$. Such equations are analogous to those found in refs.
\ref{18,20} and are rather messy.

The action of the Slavnov operator $s$ on the components fields
$\psi_\eta$, $\nu_\eta$ and $y_\eta$ can be found by substituting
$(9.16)$--$(9.18)$ into $(8.4)$--$(8.6)$ and then using $(9.14)$
to compute the Lie brackets. The resulting expressions are non local,
since they contain the $\psi_\eta$'s which, being solutions of $(8.3)$,
are non local functionals of the $\nu_\eta$'s. Further, they
depend on the $A_1$ embedding used. However, one observes the
following general structure:
$$s\hat\nu_o=\big(\bar\partial-\hat\nu_o\partial
+(\partial\hat\nu_o)\big)\hat y_o+\ldots,\eqno(9.19a)$$
$$s\nu_\eta
=\big(\bar\partial-\hat\nu_o\partial+l_\eta(\partial\hat\nu_o)\big)y_\eta
+\hat y_o\partial\nu_\eta-l_\eta\partial \hat y_o\nu_\eta+\ldots,
\quad \eta\not=o,\eqno(9.19b)$$
$$s\hat y_o=\hat y_o\partial \hat y_o+\ldots,\eqno(9.20a)$$
$$sy_\eta=\hat y_o\partial y_\eta-l_\eta\partial \hat y_oy_\eta+\ldots,
\quad \eta\not=o,\eqno(9.20b)$$
where  $\hat\nu_o=-2^{-{1\over 2}}\nu_o$ and $\hat y_o=-2^{-{1\over 2}}
y_o$ and the ellipses denote contributions of order at least one
in both the $\nu_\eta$'s and the $y_\eta$'s with $\eta\not=o$.
These relations are manifestly of the same form as those encountered
in standard $W$ gravity.

The field $\hat\nu_o$ has many of the formal properties of a Beltrami
differential $\mu$. Indeed, this is the name usually adopted in the
literature. However, such terminology is unwarranted, for $\hat\nu_o$ does
not represent any deformation of the holomorphic structure of $\Sigma$ as
appears from the fact that it does not necessarily satisfy the bound
$(2.4)$. Similarly, $\hat y_0$ shares many of the formal properties of
the diffeomorphism ghost $C$ but is distinct from it.

When $\hat A^*$ and $\hat X$ have the form $(9.17)$ and $(9.18)$, the
projective part of the automorphism anomaly
$${\cal A}_{\rm aut}(\hat X,\hat A^*;\hat A)=
{\cal A}_{\rm aut}(\hat X_0,\hat A^*{}_0;\hat A)
+{1\over 12\pi}\int_\Sigma\tr_R\big([\hat A^*,\hat X-\hat X_0]\hat B\big),
\eqno(9.21)$$
where $\hat B$, $\hat A^*$ and $\hat X$ are given by $(9.16)$--$(9.18)$
and $\hat A^*{}_0$ and $\hat X_0$ are given by $(9.17)$--$(9.18)$
with $\hat B$ set to zero. To show the above identity, one has to use
$(8.3)$, $(9.15)$, and $(9.16)$--$(9.18)$ and the fact the components
of $\hat A^*-\hat A^*{}_0$ and $\hat X-\hat X_0$ with respect to
the basis $\{t_{\eta,m_\eta}\}$ with $m_\eta=l_\eta,l_\eta-1$
vanish. The first term in the right hand side of $(9.21)$ can
be computed explicitly by using $(9.1)$, $(9.17)$ and $(9.18)$. The
following expression is found:
$${\cal A}_{\rm aut}(\hat X_0,\hat A^*{}_0;\hat A)
={-1\over 12\pi}\sum_{\eta\in\Pi}N_\eta
\bigg[\prod_{m=1}^{2l_\eta}{2\over G^{+1}{}_{l_\eta,l_\eta-m}}\bigg]
\int_\Sigma{d\bar z\wedge dz\over 2i}y_\eta D_{l_\eta}(\partial,{\cal R})
\nu_\eta,\eqno(9.22)$$
where $D_l(\partial,{\cal R})$ is the $l$--th Bol operator \ref{43}:
$$\eqalignno{
D_1(\partial,{\cal R})&=\partial^3+2{\cal R}\partial+(\partial{\cal R}),&\cr
D_2(\partial,{\cal R})&=\partial^5+10{\cal R}\partial^3+
15(\partial{\cal R})\partial^2+[9(\partial^2{\cal R})+16{\cal R}^2]
+[(\partial^3{\cal R})+{\cal R}(\partial{\cal R})],~{\rm etc.}.
\hskip1.2truecm&(9.23)\cr}$$
In general, $D_l(\partial,{\cal R})$ is of the form $\partial^{2l+1}+
O(\partial^{2l-1})$. It can be shown that there is a
projective coordinate structure $\sans p$ subordinated to
the holomorphic structure $\sans a$ such that ${\cal R}_a=0$
for $a\in{\sans p}$ \ref{35}. In such coordinates,
$D_l(\partial,{\cal R})_a=\partial_a{}^{2l+1}$ exactly.
It is remarkable that
$(9.23)$ is precisely of the form of the standard minimal $W$ anomaly
in the large central charge limit.
The term of the anomaly corresponding to the representation $\eta=o$
is formally identical to an improvement term of the diffeomorphism anomaly,
though the fields appearing are $\nu_o$ and $y_o$ rather than $\mu$
and $C$. However, upon passing form the Ward identity to the operator
product expansion formulation, this distinction is lost.
The second term in the right hand side of $(9.21)$ is a non minimal
contribution to the anomaly \ref{12}.
\vskip.4cm
$c)$ {\it The Donaldson action and Toda field theory.}
\vskip.4cm
The Drinfeld-Sokolov $DS(t,R)$ vector bundle admits a natural special
Hermitian structure $(g^\circ,\hat G)$ subordinated to $\sans A$,
where $g^\circ$ is any Hermitian metric of $\Sigma$ subordinated to
$\sans a$ and
$$\hat G=\exp(-\gamma_{g^\circ}t_{-1})\exp(-\ln g^\circ t_0)
\exp(-\bar\gamma_{g^\circ}t_{+1}),\eqno(9.24)$$
where the connection $\gamma_{g^\circ}$ is given by $(2.11)$.
Here, it is assumed that the $A_1$ embedding is such that
$t_d{}^\dagger=t_{-d}$ in order to ensure the Hermiticity of $\hat G$.
This structure is just an element of a distinguished class of special
Hermitian structures. These are of the form $(h^\circ,\hat H)$, where
$h^\circ=\exp\phi^\circ g^\circ$ is another Hermitian metric
of $\Sigma$ subordinated to $\sans a$ and
$$\hat H=\exp(-\gamma_{g^\circ}t_{-1})\exp(-\tilde\phi-\ln g^\circ t_0)
\exp(-\bar\gamma_{g^\circ}t_{+1}),\eqno(9.25)$$
with $\tilde\phi$ is Hermitian field of $S(\Sigma,1\otimes
{\sans g})$ such that $\ad t_0\tilde\phi=0$.
$\hat H$ can be expressed in terms of the traceless Donaldson
field $\hat\Phi$ and of $\hat G$.

It is interesting to compute the projective part of the Donaldson
action $S_D(\hat\Phi,0,0,0$, $0;\hat G)$. I shall present the
computation for the principal $A_1$ embedding.
Let $\Delta$ be the set of simple roots of $\sans g$,
$e_{\pm\alpha}$ and $r_\alpha$, $\alpha\in\Delta$ be the
generators of a Cartan-Weyl basis of $\sans g$ and
$C_{\alpha,\beta}$ be the Cartan matrix of $\sans g$. One has
$$[r_\alpha,r_\beta]=0,
\quad [r_\alpha,e_{\pm\beta}]=\pm C_{\alpha,\beta}e_{\pm\beta},
\quad [e_\alpha,e_{-\beta}]=\delta_{\alpha,\beta}r_\alpha,\eqno(9.26)$$
$$n_R\tr_R(r_\alpha r_\beta)=2C_{\alpha,\beta}/\beta^2,\quad
n_R\tr_R(r_\alpha e_{\pm\beta})=0,\quad
n_R\tr_R(e_\alpha e_{-\beta})=2\delta_{\alpha,\beta}/\beta^2,\eqno(9.27)$$
for $\alpha,\beta\in\Delta$, where  $n_R$ is a normalization such that
$n_R(2/\alpha^2)^2\tr_R(r_\alpha{}^2)=2$ for the long roots $\alpha$.
The generators of the principal $A_1$ are
$$t_{\pm 1}=\sum_{\alpha\in\Delta}\bigg[2\sum_{\beta\in\Delta}
C^{-1}{}_{\beta\alpha}\bigg]^{1\over 2}e_{\pm\alpha},\quad
t_0=\sum_{\alpha,\beta\in\Delta}C^{-1}{}_{\beta,\alpha}r_\alpha.\eqno(9.28)$$
$\tilde\phi$ can be expanded in terms of the coroots $r_\alpha$:
$$\tilde\phi=\sum_{\alpha\in\Delta}\phi_\alpha r_\alpha,\eqno(9.29)$$
where the $\phi_\alpha$'s are realvalued elements of $S(\Sigma,1)$.
The calculation of the Donaldson action $(7.13)$ is based on
$(9.26)$--$(9.27)$ and is totally straightforward though somewhat
lengthy. The result found is the following:
$$\eqalignno{&S_D(\hat\Phi,0,0,0,0;\hat G)=
{-1\over 12\pi n_R}\int_\Sigma{d\bar z\wedge dz\over 2i}\bigg\{
{1\over 2}\sum_{\alpha,\beta\in\Delta}{2C_{\alpha,\beta}\over\beta^2}
\bar\partial\phi_\alpha\partial\phi_\beta
-\bar\partial\partial\ln g^\circ\sum_{\alpha\in\Delta}
{2\over\alpha^2}\phi_\alpha&\cr&
+{1\over 2}g^\circ R_{g^\circ}{}^2\sum_{\beta,\gamma\in\Delta}
{2C^{-1}{}_{\beta,\gamma}\over\gamma^2}\bigg[
\exp\bigg(-\sum_{\alpha\in\Delta}\phi_\alpha C_{\alpha,\gamma}\bigg)-1\bigg]
\bigg\},&(9.30)\cr}$$
where $R_{g^\circ}$ is given by $(2.12)$.
The action $(9.30)$ is that of a Toda field theory with no exponential
term. The last term cannot be identified with the customary exponential
term of Toda field theory, since it vanishes in flat space because of
the $R_{g^\circ}{}^2$ prefactor. The reason why the exponential
interaction does not appear can be traced back to the absence of the
generalized  cosmological term in the Donaldson action.
\vskip.4cm
\par\noindent
{\bf Acknowledgements.} I wish to voice my gratitude to F. Bastianelli,
F. Ravanini, M. Stanishkov, K. Yoshida and particularly to R. Stora
for many helpful discussions.
\vskip.6cm
\centerline{\bf REFERENCES}
\def\ref#1{\lbrack #1\rbrack}
\def\NP#1{Nucl.~Phys.~{\bf #1}}
\def\PL#1{Phys.~Lett.~{\bf #1}}

\def\CMP#1{Commun.~Math.~Phys.~{\bf #1}}
\def\PR#1{Phys.~Rev.~{\bf #1}}

\def\MPL#1{Mod.~Phys.~Lett.~{\bf #1}}
\def\IJMP#1{Int.~J.~Mod.~Phys.~{\bf #1}}
\def\PREP#1{Phys.~Rep.~{\bf #1}}

\def\CQG#1{Class.~Quantum~Grav.~{\bf #1}}
\vskip.4cm
\par\noindent

\item{\ref{1}}
A. M. Polyakov, \MPL{A2} (1987) 893.

\item{\ref{2}}
V. G. Knizhnik, A. M. Polyakov and A. B. Zamolodchikov, \MPL{A3} (1988) 819.

\item{\ref{3}}
A. Alekseev and S. Shatashvili, \NP{B323} (1989) 719.

\item{\ref{4}}
M. Bershadsky and H. Ooguri, \CMP{126} (1989) 49.

\item{\ref{5}}
P. Bouwknegt and K. Schoutens, \PREP{223} (1993) 183.

\item{\ref{6}}
A. Bilal and J.-L. Gervais, \NP{B326} (1989) 222.

\item{\ref{7}}
S. R. Das, A. Dhar and K. S. Rama, preprint TIFR/TH/91-20.

\item{\ref{8}}
C. N. Pope, L. J. Romans and K. S. Stelle \PL{B269} (1991) 287.

\item{\ref{9}}
A. M. Polyakov, \IJMP{A5} (1990) 833.

\item{\ref{10}}
E. Bergshoeff, A. Bilal and K. S. Stelle \IJMP{A6} (1991) 4959.

\item{\ref{11}}
C. M. Hull, proc. of {\it Strings and Symmetries 1991}, Stony Brook,
May 1991, ed. N. Berkovits {\it et al.} (World Scientific,
Singapore, 1992).

\item{\ref{12}}
K. Schoutens. A. Sevrin and P. van Nieuwenhuizen, \PL{B243} (1990) 245,
\PL{B251} (1990) 355, \NP{B349} (1991) 791, \NP{B364} (1991) 584,
\NP{B371} (1992) 315 and proc. of {\it Strings and Symmetries 1991},
Stony Brook, May 1991, ed. N. Berkovits {\it et al.} (World Scientific,
Singapore, 1992).

\item{\ref{13}}
H. Ooguri, K. Schoutens. A. Sevrin and P. van Nieuwenhuizen,
Commun. Math. Phys. {\bf 145} (1992) 515.

\item{\ref{14}}
E. Bergshoeff, C. N. Pope and K. S. Stelle, \PL{B249} (1990) 208.

\item{\ref{15}}
G. Sotkov, M. Stanishkov and C. J. Zhu, \NP{356} (1991) 245.

\item{\ref{16}}
G. Sotkov and M. Stanishkov, \NP{B356} (1991) 439.

\item{\ref{17}}
A. Gerasimov, A. Levin and A. Marshakov, \NP{B360} (1991) 537.

\item{\ref{18}}
A. Bilal, V. V. Fock and I. I. Kogan, \NP{B359} (1991) 635.

\item{\ref{19}}
K.Yoshida, \IJMP{A7} (1992) 4353.

\item{\ref{20}}
R. Zucchini, \CQG{10} (1993) 253.

\item{\ref{21}}
C. M. Hull, \PL{B269} (1991) 257 and preprints LANL hep-th/9211113 and
LANL hep-th/9303071.

\item{\ref{22}}
J.-L. Gervais and Y. Matsuo, \PL{B274} (1992) 309 and
Commun. Math. Phys. {\bf 152} (1993) 317.

\item{\ref{23}}
J. De Boer and J. Goeree, \PL{B274} (1992) 289 and
preprint LANL hep-th/9206098.

\item{\ref{24}}
A. A. Belavin and V. G. Knizhnik, \PL{B168} (1986), 201.

\item{\ref{25}}
M. Knecht, S. Lazzarini and R. Stora, \PL{B262} (1991) 25.

\item{\ref{26}}
M. Knecht, S. Lazzarini and R. Stora, Phys. Lett. {\bf B273} (1991) 63.

\item{\ref{27}}
H. Verlinde, \NP{B337} (1990) 652.

\item{\ref{28}}
M. Knecht, S. Lazzarini and F. Thuillier, \PL{B251} (1990), 279.

\item{\ref{29}}
S. K. Donaldson, Proc. London Math. Soc. {\bf 50} (1985) 1 and
Duke Math. J. {\bf 54} (1987) 231.

\item{\ref{30}}
F. David, \MPL{A3} (1988) 1651.

\item{\ref{31}}
J. Distler and H. Kaway, \NP{B321} (1989) 509.

\item{\ref{32}}
E. Witten, \PR{D44} (1991) 314.

\item{\ref{33}}
A. Sevrin, K. Thielemans and W. Troost, preprints LANL hep-th/9303133,
and LANL hep-th/9306033.

\item{\ref{34}}
A. Sevrin, LANL hep-th/9306059.

\item{\ref{35}}
R. Gunning, {\it Lectures on Riemann surfaces}, Princeton University Press
(1966) and references therein.

\item{\ref{36}}
R. Gunning, {\it Lectures on Vector Bundles on Riemann surfaces}, Princeton
University Press (1967) and references therein.

\item{\ref{37}}
S. Lazzarini, Doctoral thesis, LAPP Annecy-le-Vieux, (1990) and references
therein.

\item{\ref{38}}
A. M. Polyakov, \PL{B103} (1981) 207 and \PL{B103} (1981) 211.

\item{\ref{39}}
R. Zucchini, \PL{B260} (1991) 296 and \CMP{152} (1993) 269.

\item{\ref{40}}
S. Kobayashi and K. Nomizu, {\it Foundations of Differential Geometry},
vols. I and II, J. Wiley \& Sons (1963).

\item{\ref{41}}
R. O. Wells, {\it Differential Analysis on Complex Manifolds},
Graduate Texts in Mathematics, Springer Verlag (1980) and references
therein.

\item{\ref{42}}
V. G. Drinfeld and V. V. Sokolov, J. Sov. Math. {\bf 30} (1985) 1975.

\item{\ref{43}}
F. Gieres, \IJMP{A8} (1993) 1.

\bye